\documentclass[12pt]{article}
\pdfoutput=1

\usepackage{amsmath}
\usepackage{amssymb}
\usepackage{graphicx}
\usepackage{cite}
\usepackage{multirow}
\usepackage{longtable}
\usepackage{lscape}
\usepackage{subfigure}

\newcommand{\beq}{\begin{equation}}
\newcommand{\eeq}{\end{equation}}
\newcommand{\be}{\begin{eqnarray}}
\newcommand{\ee}{\end{eqnarray}}

\newcommand{\ms}{\Delta m^2_{21}}
\newcommand{\ma}{\Delta m^2_{31}}

\newcommand{\stch}{\sin^2 2\theta_{13}}

\newcommand{\mst}{\Delta m^2_{21}{\mbox {(true)}}}
\newcommand{\mat}{\Delta m^2_{31}{\mbox {(true)}}}

\newcommand{\stcht}{\sin^2 2\theta_{13}{\mbox {(true)}}}

\newcommand{\dcpt}{\delta_{CP}{\mbox {(true)}}}
\newcommand{\dcp}{\delta_{\mathrm{CP}}}
\newcommand{\tmt}{\theta_{23}}
\newcommand{\tet}{\theta_{13}}
\newcommand{\tem}{\theta_{12}}

\def\nue{{\nu_e}}
\def\anue{{\bar\nu_e}}
\def\numu{{\nu_{\mu}}}
\def\anumu{{\bar\nu_{\mu}}}
\def\nutau{{\nu_{\tau}}}

\newcommand{\capdef}{}
\newcommand{\mycaption}[2][\capdef]{\renewcommand{\capdef}{#2}
       \caption[#1]{{\footnotesize #2}}}
\makeatletter
\renewcommand{\fnum@table}{\textbf{\tablename~\thetable}}
\renewcommand{\fnum@figure}{\textbf{\figurename~\thefigure}}
\makeatother

\begin{document}

\begin{titlepage}

\renewcommand{\thefootnote}{\alph{footnote}}

\vspace*{-3.cm}
\begin{flushright}
EURONU-WP6-11-38  \\
IFIC/11-48
\end{flushright}

\vspace*{0.2cm}

\renewcommand{\thefootnote}{\fnsymbol{footnote}}
\setcounter{footnote}{-1}

{\begin{center} 

{\large\bf

An incremental approach to unravel the neutrino mass hierarchy and CP violation with a long-baseline Superbeam for 
large $\theta_{13}$ \\[0.3cm]

}

\end{center}} 

\renewcommand{\thefootnote}{\alph{footnote}}

\vspace*{.6cm}

{\begin{center} {
               \large{\sc
                 Sanjib Kumar Agarwalla\footnote[1]{\makebox[1.cm]{Email:}
                 Sanjib.Agarwalla@ific.uv.es}},
               \large{\sc
                 Tracey Li\footnote[2]{\makebox[1.cm]{Email:}
                 tracey.li@ific.uv.es}},
               \large{\sc
                 Andr{\' e} Rubbia\footnote[3]{\makebox[1.cm]{Email:}
                 andre.rubbia@cern.ch}}
                 }
\end{center}
}
\vspace*{0cm}
{\it
\begin{center}

\footnotemark[1]${}^,$\footnotemark[2]
       Instituto de F\'{\i}sica Corpuscular, CSIC-Universitat de Val\`encia, \\
       Apartado de Correos 22085, E-46071 Valencia, Spain \\
\footnotemark[3]
       ETH Zurich, Institute for Particle Physics, CH-8093 Z{\"u}rich, Switzerland
\end{center}}

\vspace*{0.5cm}

{\Large \bf
\begin{center} Abstract \end{center}  }

Recent data from long-baseline neutrino oscillation experiments have provided new information on $\theta_{13}$,
hinting that $0.01\lesssim\stch \lesssim 0.1$ at $2\,\sigma$ confidence level. In the near future, further confirmation of this result 
with high significance will have a crucial impact on the optimization of the future long-baseline neutrino oscillation experiments 
designed to probe the neutrino mass ordering and leptonic CP violation.
In this context, we expound in detail the physics reach of an experimental setup where neutrinos produced in a conventional wide-band beam  
facility at CERN are observed in a proposed Giant Liquid Argon detector at the Pyh\"asalmi mine, at a distance of $2290\,\mathrm{km}$. 
Due to the strong matter effects and the high detection efficiency at both the first and second oscillation maxima, this particular setup would have
unprecedented sensitivity to the neutrino mass ordering and leptonic CP violation in the light of the emerging hints of large $\theta_{13}$.
With a 10 to $20\,\mathrm{kt}$ `pilot' detector and just a few years of neutrino beam running, the neutrino mass hierarchy could be determined, 
irrespective of the true values of $\dcp$ and the mass hierarchy, at $3\,\sigma$ ($5\,\sigma$) confidence level if $\stcht$ = 0.05 (0.1).
With the same exposure, we start to have $3\,\sigma$ sensitivity to CP violation if $\stcht > 0.05$, in particular testing maximally 
CP-violating scenarios at a high confidence level. After optimizing the neutrino and anti-neutrino running fractions, we study the performance of the setup 
as a function of the exposure, identifying three milestones to have roughly 30\%, 50\% and 70\% coverage in $\dcpt$ for $3\,\sigma$ CP violation discovery.
For comparison, we also study the CERN to Slanic baseline of $1540\,\mathrm{km}$.
This work nicely demonstrates that an incremental program, staged in terms of the exposure, can achieve the desired physics goals 
within a realistically feasible timescale, and produce significant new results at each stage. 

\vspace*{.5cm}

\end{titlepage}
\newpage

\renewcommand{\thefootnote}{\arabic{footnote}}
\setcounter{footnote}{0}

\section{Introduction and motivation}
\label{sec:intro-motiv}

Over the last thirteen years or so, marvellous data from world-class experiments involving neutrinos from 
the sun~\cite{Cleveland:1998nv,Altmann:2005ix,Hosaka:2005um,Ahmad:2002jz,Aharmim:2008kc,Aharmim:2009gd,Arpesella:2008mt},
the Earth's atmosphere~\cite{Fukuda:1998mi,Ashie:2005ik}, nuclear reactors~\cite{Araki:2004mb,:2008ee}, and 
accelerators~\cite{Ahn:2006zza,Adamson:2008zt} have firmly established the phenomenon of neutrino oscillations. This implies that neutrinos have mass, an exclusive example of experimental evidence for physics beyond
the Standard Model. The T2K experiment in Japan is a new addition to this remarkable list of successful experiments and has recently 
provided an important breakthrough in the $\theta_{13}$-driven $\numu \to \nue$ appearance oscillation channel.

The T2K experiment has reported the first indication of electron neutrino appearance in the Super-Kamiokande detector, located at a distance of 295 km from an accelerator-based $\numu$ source. The neutrinos have an energy of roughly 0.6 GeV and are produced at J-PARC~\cite{Abe:2011sj}. The observation of six electron-like events with an estimated background of 1.5 events is consistent with subdominant $\nu_{\mu} \to \nu_e$ 
transitions in the atmospheric sector and indicates a non-zero value of $\theta_{13}$ at $2.5\,\sigma$ significance. 
Within a couple of weeks of the T2K results, the MINOS collaboration announced the observation of 62 electron-like events with 
an estimated background of 49 events~\cite{Adamson:2011qu}, which excludes a zero $\theta_{13}$ at 89\% C.L. ($1.7\,\sigma$).

These new results have now been included in global fits of all available neutrino oscillation data~\cite{Fogli:2011qn,Schwetz:2011zk}, 
which point towards a non-zero value of $\theta_{13}$ at around $3\,\sigma$ C.L. and hint that $\sin^22\theta_{13}$ lies between 
0.01 and 0.1 at $2\,\sigma$ C.L. with a best-fit value of 0.051 (0.063) for normal (inverted) mass ordering~\cite{Schwetz:2011zk}.

With further running of the T2K~\cite{Oyama:2011pt} and MINOS~\cite{Adamson:2011qu} experiments, and the impressive 
start-up of reactor neutrino experiments (Double Chooz~\cite{Kuze:2011ic}, RENO~\cite{Ahn:2010vy}, and Daya Bay~\cite{Lin:2011pb})
and of the upcoming accelerator experiment NO$\nu$A~\cite{Ayres:2004js}, it is likely that the positive evidence for $\theta_{13}$ 
will be confirmed with high statistical significance in the next few years. A non-zero, and in particular a large value of $\theta_{13}$ will 
provide a boost to the field of neutrino oscillation physics, making it possible to discover leptonic CP violation (CPV) if the Dirac CP phase, 
$\dcp$, is not equal to $0^{\circ}$ or $180^{\circ}$, and allowing for a direct determination of $sgn(\ma)$ {\it aka}, the neutrino mass 
hierarchy\footnote{In this paper we define $\Delta m^2_{ij} = m_i^2 - m_j^2$ and refer to $sgn(\ma)$ as the neutrino mass 
hierarchy -- $sgn(\ma) >0$ is called `normal hierarchy' (NH) while $sgn(\ma) <0$ is called `inverted hierarchy' (IH). The arguments 
are valid for both hierarchical as well as quasi-degenerate neutrino mass spectra.}. 

Given their relatively short baselines, narrow band beams and limited statistics, the present generation T2K and NO$\nu$A experiments 
have a limited reach in probing CPV and the neutrino mass hierarchy~\cite{Huber:2009cw}, even for the large values of $\tet$ indicated 
by the recent global fits~\cite{Fogli:2011qn,Schwetz:2011zk}. Hence, new long-baseline experiments with intense 
neutrino beam sources and advanced detector technologies are needed~\cite{Bandyopadhyay:2007kx} to fathom the hitherto uncharted parameter 
space of the neutrino mixing matrix well beyond the capabilities of T2K and NO$\nu$A.

As discussed in the literature (see e.g.~Refs.~\cite{Bandyopadhyay:2007kx,Mezzetto:2010zi}), the ability of future long-baseline neutrino experiments 
to discover CPV and the mass hierarchy depends on the achievable event statistics and hence strongly on the value of $\theta_{13}$. 
The definitive confirmation of large $\theta_{13}$ will have a significant impact on the realization of future long-baseline neutrino oscillation 
experiments, for which planning has up to now focused on a staged approach to achieve sensitivity to increasingly smaller values of $\theta_{13}$. 
This approach has been exemplified in the optimization of the Neutrino Factory~\cite{Agarwalla:2010hk} and 
Beta-beam~\cite{Agarwalla:2006vf,Agarwalla:2008gf} for which 
it is possible to discover a value of $\sin^22\theta_{13}$ as small as $10^{-4}$. 
However, following the evidence for a relatively large value of $\theta_{13}$, the focus of future optimizations will be the possibility to explore CPV and the 
mass hierarchy for a {\it given} value of $\stch$.

Accordingly, in this paper, we analyze the impact of large $\theta_{13}$ on the optimization of future long-baseline superbeam 
experiments with a special emphasis on potential European scenarios~\cite{Autiero:2007zj,Rubbia:2010fm,Angus:2010sz,Rubbia:2010zz}. 
The technology of conventional beam is well established and understood, and when capable of sustaining very high 
proton beam powers (above a megawatt), they are known as superbeams. Superbeam experiments are `super' in the sense that they use proton beams of 
unprecedented strength and detectors with huge fiducial masses (tens or hundreds of kt). 
In a superbeam, we shoot a target with as many protons as can 
be provided, and then the produced mesons (mostly pions and some kaons) are focused and sign-selected using a magnetic horn. 
Henceforth, we will denote the beam produced by positive horn focusing as `PHF' and the beam created by negative horn focusing as 
`NHF'. In our study, we rely on a wide-band beam (WBB) approach, a type of superbeam where most of the produced pions are utilized and the on-axis detector 
is exposed to a wide range of energies. This results in higher statistics in comparison with an off-axis or a narrow-band beam and also allows several $L/E$ values 
to be studied in a single experiment, which helps to resolve parameter degeneracies~\cite{Huber:2010dx}.
The main caveat with the WBB approach is the presence of a high-energy tail in the beam which enhances the feed-down of neutral current (NC) 
background rates. Therefore, an advanced detector technology with good energy resolution and good NC rejection capability is vital to use the 
broad spectrum efficiently. 

In this paper, we focus on the CERN to Pyh{\"a}salmi baseline of $2290\,\mathrm{km}$\footnote{Note that this baseline is very close
to the `bimagic' distance~\cite{Raut:2009jj,Dighe:2010js}.} which is the longest baseline considered in the 
LAGUNA design study~\cite{Angus:2010sz,Rubbia:2010zz}. An outcome of this study is that the Pyh{\"a}salmi mine in Finland has the appropriate
infrastructure, at the right depth, to house an underground $100\,\mathrm{kt}$ liquid argon detector~\cite{Nuijten:2011zz}.
The baseline is also sufficiently long such that matter effects are quite strong and we can expect that the mass hierarchy can be determined at a high 
confidence level if $\theta_{13}$ is large, irrespective of the true values\footnote{We distinguish between the `true' values of the oscillation parameters, which are the values chosen by Nature (or the values we assume that Nature has chosen and which we use to compute the data), and their fitted values. Throughout this paper we indicate the true value of a parameter by writing `(true)' after that parameter.} of the other oscillation parameters, much earlier than before we start to have sensitivity to CPV. Additionally, the first and second oscillation maxima for the CERN-Pyh{\"a}salmi baseline are at $4.5\,\mathrm{GeV}$ 
and $1.5\,\mathrm{GeV}$ for $\ma = 2.45 \times 10^{-3}\,\mathrm{eV}^2$, well above the typical threshold of a detector\footnote{Note that the first oscillation maximum is also above the energy of the tau kinematic threshold. This class of events was studied in Ref.~\cite{Bueno:2004ty}.}. This enables us to observe the first and second oscillation 
maxima with high efficiency, and also the uncertainties on cross-sections due to nuclear effects are negligible at these energies. Higher energies are also preferred experimentally as interactions are easier to identify and the total cross-section increases with energy.

We compare the physics reach of the CERN to Pyh{\"a}salmi baseline with the CERN to Slanic\footnote{The Slanic site in Romania provides in practice only 
a shallow depth, which would greatly limit the sensitivity to astrophysical sources and proton decay.} baseline of $1540\,\mathrm{km}$, the 
second longest baseline option in Europe. This baseline is of a fairly similar distance to the Fermilab to Homestake baseline of $1290\,\mathrm{km}$, being actively studied for the
Long-Baseline Neutrino Experiment (LBNE)~\cite{Diwan:2003bp,Barger:2007yw,Huber:2010dx} in the USA. We present a comparison between the CERN-Pyh{\"a}salmi and CERN-Slanic baselines, in terms of sensitivity to the mass hierarchy and CPV.    

As far as the choice of far detector\footnote{In the real experiment, we will also have a small near detector to reduce the impact of
systematic uncertainties.} is concerned, here we will focus on a Giant Liquid Argon Charge Imaging Experiment with a target mass
up to $100\,\mathrm{kt}$, as proposed in Refs.~\cite{Rubbia:2004tz,Rubbia:2009md}. An initial `pilot' version could have a mass of 10 to $20\,\mathrm{kt}$. 
The Liquid Argon Time Projection Chamber (LArTPC) technique~\cite{Amerio:2004ze} offers an excellent environment in which to reconstruct electron 
events in the GeV range and above with high efficiency, while considerably suppressing the NC background. 

A relatively large value of $\tet$ allows us to pursue a staged approach in terms of the size of the experiment. For a given experimental
setup, the physics sensitivity is mainly dictated by the incident proton beam power on the target~\cite{Ferrari:2002yj}, and by the fiducial 
mass of the far detector. In this study, we assume a CERN-based scenario with a primary proton 
energy of $50\,\mathrm{GeV}$\footnote{We plan to perform a similar study with a primary proton beam energy of $400\,\mathrm{GeV}$.} ~\cite{Rubbia:2010fm} which delivers $3 \times 10^{21}$ protons on target (pot) per year, using a beam power of $1.6\,\mathrm{MW}$ and running for 200 days per calendar year. In order to demonstrate our staged experimental strategy, we introduce the quantity `exposure' which is defined as
\beq
{\rm exposure} =  ({\rm pot\,per\,year}) \times ({\rm fiducial\,mass\,of\,detector\,in\,kt}) \times ({\rm total\,runtime\,in\,years}),
\label{eq:exposure}
\eeq
and has units of pot$\cdot$kt. The maximum detector mass of $100\,\mathrm{kt}$ combined with a total ten years of running (combined PHF and NHF) gives a total 
exposure of $3000 \times 10^{21}$ pot$\cdot$kt. We treat the exposure as a free parameter in our study with an upper limit of $3 \times 10^{24}$ pot$\cdot$kt and
quantify the performance of both baselines in terms of the exposure. The following are the main topics that we address in our study, 
in the light of the emerging hints of large $\tet$:

\begin{itemize}

\item
Since the determination of $sgn(\ma)$ can be accomplished with a smaller exposure than that which is needed for substantial 
CPV discovery, we first optimize the sharing of the total exposure between PHF and NHF in order to gain a $3\,\sigma$ and $5\,\sigma$ confirmation of
the mass hierarchy, for both choices of the true hierarchy and for 100\% of the values of $\dcpt$. The information obtained about $sgn(\ma)$ at the early stage 
of the experiment plays a vital role in maximizing the sensitivity of the setup to CPV - the right choice of the mass hierarchy is essentially a mandatory input 
for discovering CPV. Possibilities to determine it with atmospheric neutrinos~\cite{Petcov:2005rv,Samanta:2010xm} have been explored, but are less efficient 
and more prone to systematic errors than the direct method considered here.  

\item
We propose an incremental approach by identifying milestones of exposure which will enable the value of $\dcp$ to be gradually constrained, by incrementally increasing the CPV discovery coverage from $30\%$ to $50\%$ and finally $70\%$ of the values of $\dcpt$, at a confidence level of $3\,\sigma$. We believe that this approach is fully justified by the fact that at present, the favorable values (e.g.~close to maximal) of $\dcpt$ are still allowed~\cite{Fogli:2011qn,Schwetz:2011zk}. If Nature is as kind as it appears to be for $\tet$ such that $\dcpt$ is not close to CP conserving values $0^{\circ}$ or $180^{\circ}$, then we will be able to discover the CPV quite early in the time frame. An incremental approach 
is also justified in view of the challenges (some of them are unknown) involved in operating very high power superbeams and in building giant underground neutrino detectors, which makes such an approach effectively safer and possibly more cost-effective.

\end{itemize}

The paper is organized as follows. We begin with the relevant oscillation probabilities in Section 2.
In Section 3, we give the details of the experimental setup. We briefly describe the CERN-based superbeam facility and the fluxes 
that we use for the Pyh{\"a}salmi and Slanic baselines. Then we discuss the characteristics of the LArTPC. In Section 4 we present the expected event rates for the CERN-Pyh{\"a}salmi and CERN-Slanic baselines. The details of our numerical technique and analysis procedure are discussed in Section 5. In Section 6, we present our results. 
Finally, we summarize and draw our conclusions in Section 7.    

\section{The $\mathbf{P_{\mu e}}$ oscillation channel}
\label{sec:prob}

The use of an appearance channel in which the neutrino changes flavor between production and detection is necessary in order to probe CPV in neutrino oscillations. 
In a superbeam, we deal with the transition probabilities $P(\numu \to \nue)$ and $P(\anumu \to \anue)$ in order to explore the neutrino mass ordering and CPV.
For both the baselines considered in this work, matter effects are significant and the exact expressions of the three-flavor oscillation probabilities including matter effects 
are very complicated. Therefore, to demonstrate the nature of neutrino oscillations as a function of baseline and/or neutrino energy, 
it is useful to have an approximate analytic expression for $P_{\mu e}$ (the T-conjugate of  $P_{e \mu}$) in
matter~\cite{Wolfenstein:1977ue,Mikheev:1986gs,Barger:1980tf}, keeping terms only up to second order in
the small quantities $\theta_{13}$ and $\alpha \equiv \ms/\ma$~\cite{Cervera:2000kp,Freund:2001ui}:
\be
P_{\mu e} &\simeq&
{\underbrace{\sin^2\theta_{23} \sin^22\theta_{13}
    \frac{\sin^2[(1-\hat{A})\Delta]}{(1-\hat{A})^2}}_{C_0}}
    + {\underbrace{\alpha^2 \cos^2\theta_{23} \sin^22\theta_{12}
    \frac{\sin^2(\hat{A}\Delta)}{{\hat{A}}^2}}_{C_1}} \nonumber \\
&\mp& {\underbrace{\alpha \sin2\theta_{13}\cos\theta_{13} \sin2\theta_{12}
    \sin2\theta_{23} \sin(\Delta) \frac{\sin(\hat{A}\Delta)}{\hat{A}}
    \frac{\sin[(1-\hat{A})\Delta]}{(1-\hat{A})}}_{C_-}} \sin\dcp \nonumber \\
&+& {\underbrace{\alpha \sin2\theta_{13}\cos\theta_{13} \sin2\theta_{12}
    \sin2\theta_{23} \cos(\Delta) \frac{\sin(\hat{A}\Delta)}{\hat{A}}
    \frac{\sin[(1-\hat{A})\Delta]}{(1-\hat{A})}}_{C_+}} \cos\dcp ,
\label{eq:pmue}
\ee
where
\be
\Delta\equiv \frac{\ma L}{4E},
~~
\hat{A} \equiv \frac{A}{\ma},
~~
 A=\pm 2\sqrt{2}G_FN_eE.
\label{eq:matt}
\ee
Eq.~\eqref{eq:pmue} has been derived under the constant matter density approximation. Here, $A$ is the matter potential, 
expressed in terms of the electron density, $N_e$, and the (anti-)neutrino energy $E$ and is positive for neutrinos and negative for anti-neutrinos. 
The `$-$' sign which precedes the term $C_{-}$ refers to neutrinos whereas the `$+$' refers to to anti-neutrinos.
The first term of Eq.~\eqref{eq:pmue} ($C_0$) can be used to obtain information about the value of $\theta_{13}$. This is also the
term which contains the largest Earth matter effect and which can therefore be used to measure the sign of $\ma$. The term $C_1$ is independent of both $\theta_{13}$ and $\delta_{CP}$ and depends mainly on the solar parameters, $\ms$ and $\theta_{12}$. The term $C_{-}$ is the CP-violating part. The term $C_{+}$, although $\delta_{CP}$-dependent, is 
CP-conserving. While we will use this formula to discuss our results in some cases, all the simulations presented in this 
paper are based on the full three-flavor neutrino oscillation probabilities in matter, using the Preliminary Reference Earth Model (PREM)~\cite{Dziewonski:1981xy} 
for the Earth matter density.

In spite of using both neutrinos and anti-neutrinos, a serious complication for all long-baseline experiments involving these channels emerges from discrete degeneracies which appear in three forms: 
the ($\theta_{13},\dcp$) intrinsic degeneracy~\cite{BurguetCastell:2001ez}, the ($sgn(\ma),\dcp$) degeneracy~\cite{Minakata:2001qm}, 
and the ($\theta_{23},\pi/2-\theta_{23}$) degeneracy~\cite{Fogli:1996pv}. This leads to an eight-fold degeneracy~\cite{Barger:2001yr}, 
with several degenerate solutions in addition to the true one. The presence of these degenerate solutions can severely reduce the 
sensitivity of an experiment. For both the $2290\,\mathrm{km}$ and $1540\,\mathrm{km}$ baselines, we can extract information 
from the first as well as the second oscillation maximum quite efficiently using the LArTPC detector, and this helps to evade the problem
of intrinsic~\cite{BurguetCastell:2001ez} and $sgn(\ma)$~\cite{Minakata:2001qm} ambiguities which can cause the 
$\pi$-transit~\cite{Huber:2002mx} effect even for large values of $\stcht$.

\section{Description of experimental setup}
\label{sec:setup}

In this Section, we first present the CERN-based superbeam fluxes. Then we describe various properties of the LArTPC detector which 
are relevant to this work and summarize them in Table~\ref{tab:detector}. 

\subsection{CERN-based superbeam fluxes}
\label{sec:flux}

\begin{figure}[tp]
\centering
\includegraphics[width=0.49\textwidth]{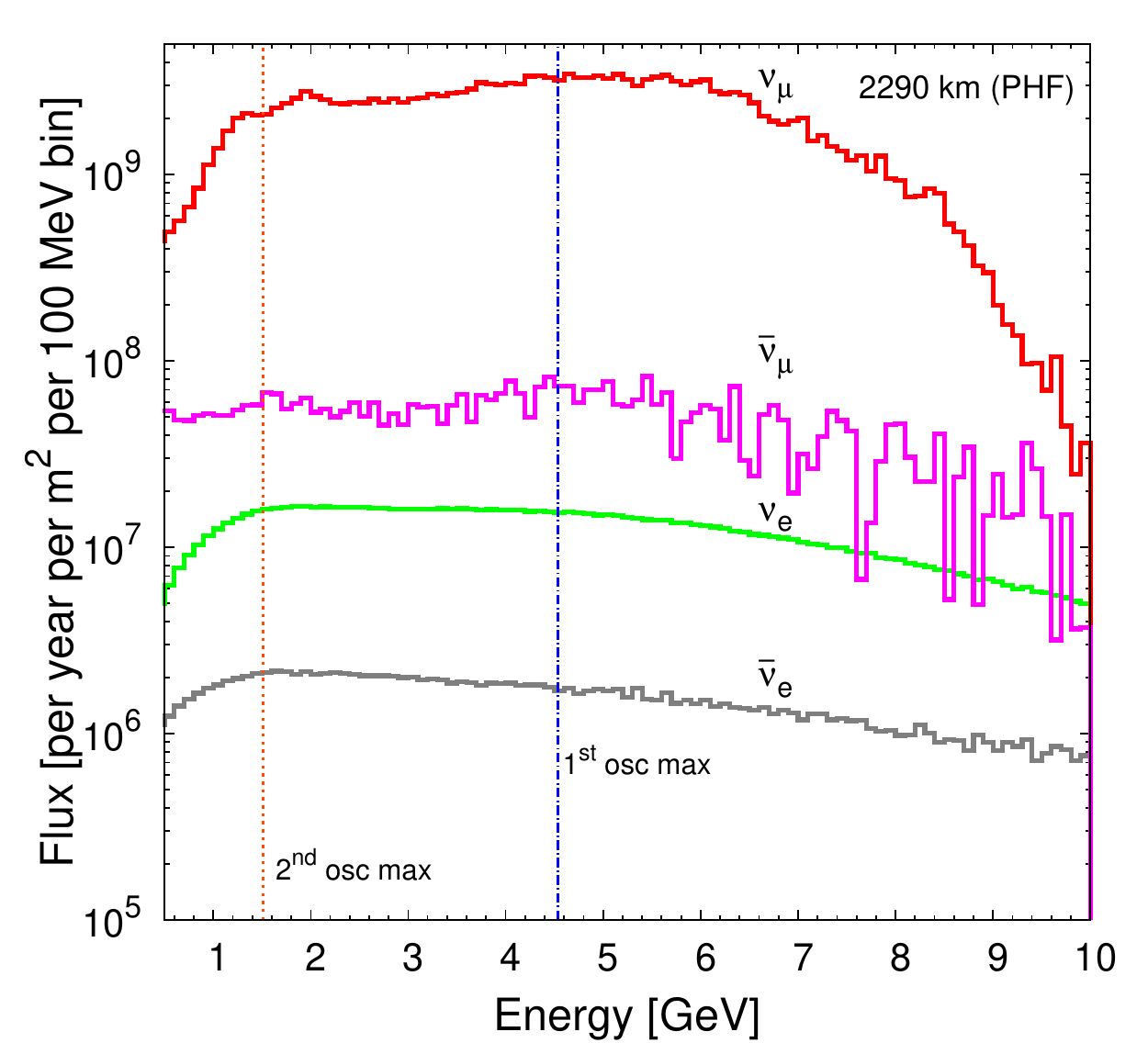}
\includegraphics[width=0.49\textwidth]{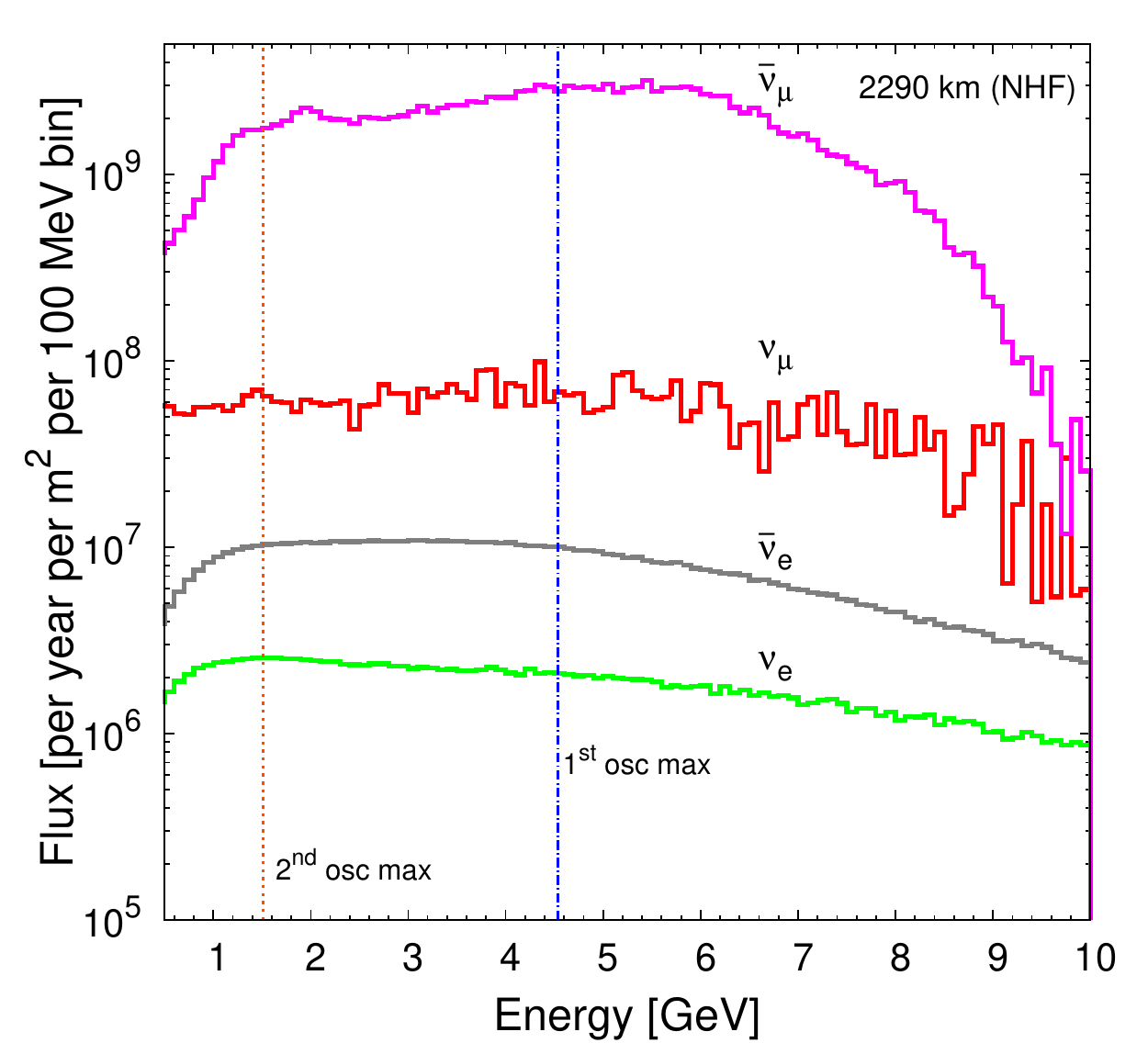}
\includegraphics[width=0.49\textwidth]{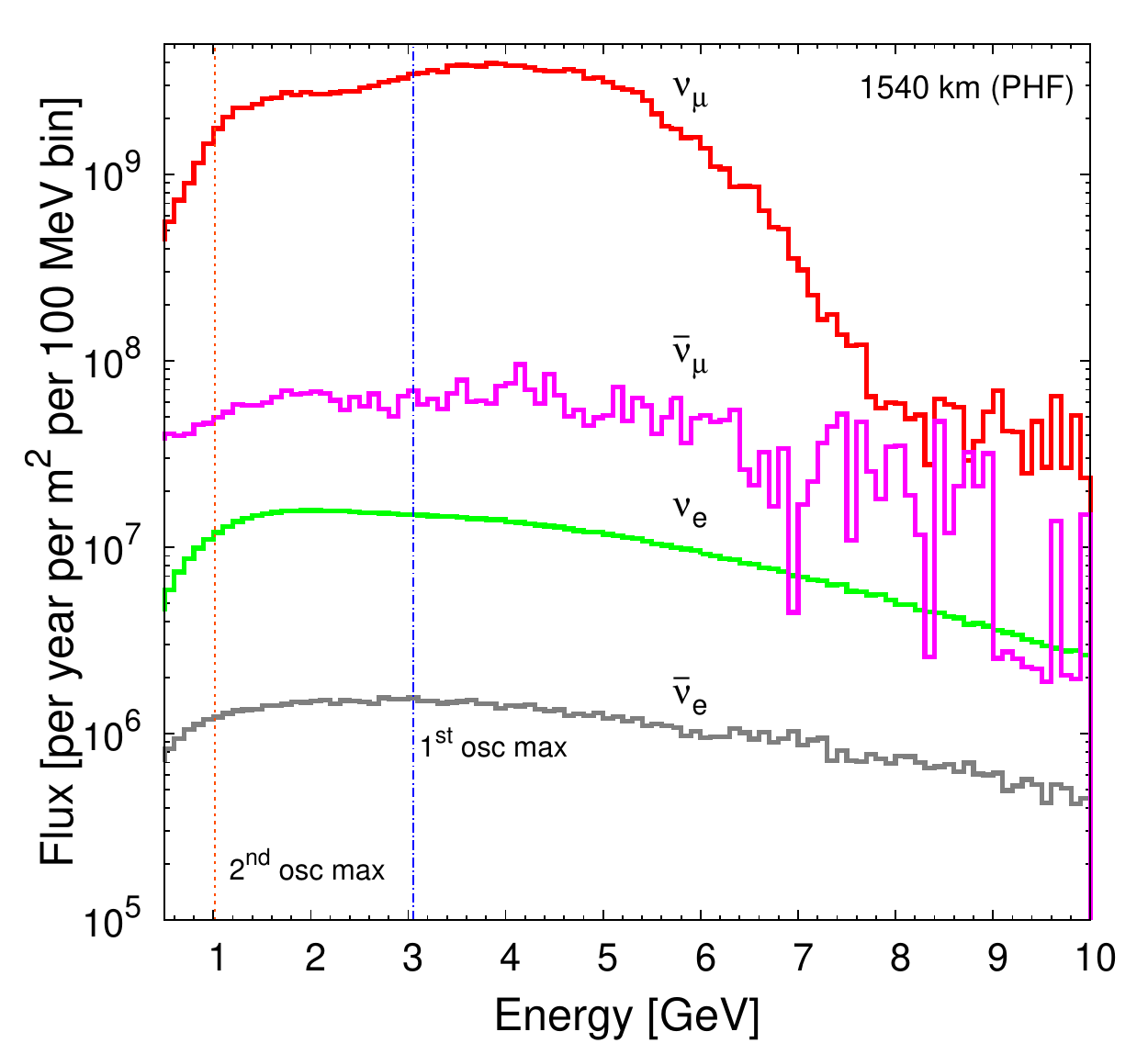}
\includegraphics[width=0.49\textwidth]{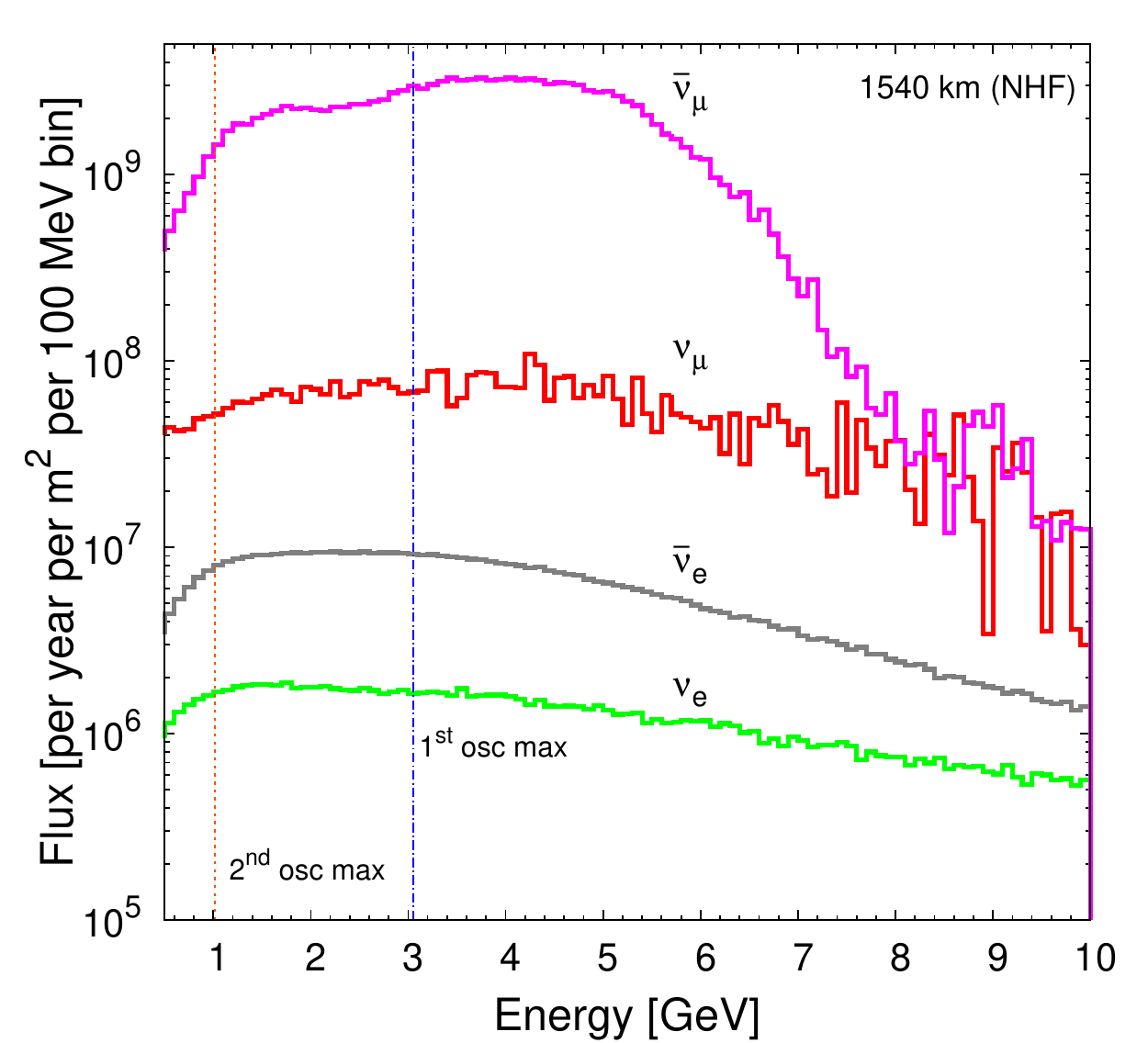}
\mycaption{\label{fig:flux} The upper panels depict the un-oscillated PHF (left panel) and NHF (right panel) superbeam flux spectrum 
arriving at a detector placed at Pyh{\"a}salmi (a distance of $2290\,\mathrm{km}$ from CERN). The lower panels show the same for the
CERN-Slanic baseline (a distance of $1540\,\mathrm{km}$). The blue dot-dashed and the orange dotted vertical
lines display the locations of the first and second oscillation maxima.}
\end{figure}

A conventional (dominantly muon) neutrino beam is produced from the decay of secondary hadrons (pions, kaons)
which are created by high-energy primary protons impinging on a long target. The neutrino beam direction is defined by the orientation
of the decay tunnel downstream of the target. An important feature is the insertion of focusing elements (so called horns), whose shape and 
current density can be chosen to direct emerging secondary hadrons of a given momentum into the decay tunnel, thereby 
defining the energy spectrum of the resulting muon neutrino beam. The magnetic property of the horn is used to select the sign of the
produced mesons. When the horn acts in the positive (negative) mode, then the resulting beam is known as `PHF' (`NHF'). 
The customary approach for long-baseline oscillation 
experiments is to tune the muon neutrino flux around the energy of the first oscillation maximum for a given baseline. For our studies, which give 
priority to the measurement of the oscillatory pattern of the flavor conversion processes as a function of the neutrino energy,
a wide-band neutrino beam tuned at the first oscillation maximum and broad enough to provide coverage of the second maximum 
is the adequate solution.

Presently the CERN to Gran Sasso Laboratory neutrino beam (CNGS)~\cite{Acquistapace:1998rv,Baldy:1999dc} is produced 
using $400\,\mathrm{GeV}$ protons from the SPS accelerator and the neutrino flux is peaked around $17\,\mathrm{GeV}$ to 
optimize the rate of $\nutau$ CC events in the far detector.

Let us now consider options for the CERN-Pyh\"asalmi and CERN-Slanic baselines. Prospects for high-intensity proton sources at 
CERN and their relevance for high intensity neutrino beams were discussed in Ref.~\cite{Rubbia:2010fm}.
The existing $400\,\mathrm{GeV}$ SPS accelerator, profiting from the LHC luminosity upgrades, will be able to deliver beams with higher 
power in the near future, but reaching megawatt beam power at the SPS will require significant replacement of its
injector chains and a stretch of its performance. A more appropriate solution which was advocated is to consider a new high power 
accelerator (HP-PS2), with a proton energy in the range $30$ to $50\,\mathrm{GeV}$~\cite{Rubbia:2010fm}, with for example
an intensity of $2.5 \times 10^{14}$ protons and a fast cycle of 1.2 seconds, corresponding to a proton beam power of $1.6\,\mathrm{MW}$. 
A conceptual design is being developed in the context of the LAGUNA-LBNO design study~\cite{Angus:2010sz,Rubbia:2010zz}.

In this paper, we have assumed that a $50\,\mathrm{GeV}$ HP-PS2 drives the neutrino beam for 200 days per calendar year, 
corresponding to an integrated $3 \times 10^{21}$ pot per year. The neutrino fluxes were computed by 
A. Longhin~\cite{Longhin:2010zz,longhin}. Also, these fluxes have been optimized for each baseline to have a peak 
at the first oscillation maximum as can be seen from Fig.~\ref{fig:flux}. The blue dot-dashed vertical line in each panel of Fig.~\ref{fig:flux} 
shows the location of first oscillation maximum which is at $3\,\mathrm{GeV}$ ($4.5\,\mathrm{GeV}$) for the CERN-Slanic (CERN-Pyh{\"a}salmi) baseline
for $\ma = 2.45 \times 10^{-3}\,\mathrm{eV}^2$. The left panels of Fig.~\ref{fig:flux} display the flux spectrum produced by PHF 
where $\numu$ is the dominant beam component and the right panels show the flux spectrum obtained from NHF where $\anumu$ dominates.

The superbeam fluxes suffer from a hurdle of an intrinsic $\nue$ ($\anue$) contamination, which poses a serious problem in the form 
of backgrounds in $\numu \to \nue$ ($\anumu \to \anue$) appearance study. At the flux level, we have 0.62\% (0.45\%) intrinsic contamination of 
$\nue$ ($\anue$) compared to $\numu$ ($\anumu$) in the beam for CERN-Pyh{\"a}salmi baseline. In the case of CERN-Slanic baseline,
the $\nue$ ($\anue$) contamination is at the level of 0.57\% (0.39\%) compared to $\numu$ ($\anumu$) in the beam. 

\subsection{The LArTPC detector}
\label{sec:LArTPC}

\begin{table}[t]
\begin{center}

\begin{tabular}{||c|c|c||} \hline\hline

\multicolumn{1}{||c||}{{\rule[0mm]{0mm}{6mm}{Detector characteristics}}}
& \multicolumn{1}{|c||}{\rule[-3mm]{0mm}{6mm}{LArTPC (Both $\mu^{\pm}$ and $e^{\pm}$)}} \cr
\hline\hline
Fiducial mass & $10 - 100\,\mathrm{kt}$ \cr
\hline
Neutrino energy threshold & $0.5\,\mathrm{GeV}$ \cr
\hline
\multirow{2}{*}{Detection efficiency ($\epsilon$)} & 100\% for $\mu^{\pm}$ \cr
& 80\% for $e^{\pm}$ \cr
\hline
Energy resolution ($\delta E$) (GeV) & 0.15$\sqrt{\rm E/GeV}$ for CC $\mu^{\pm}$ and $e^{\pm}$ sample \cr
\hline
NC background smearing & Migration matrices (different for $\nu$ and $\bar\nu$) \cr
\hline
Bin size & 0.125 GeV \cr
\hline
NC background rejection efficiency & 99.5\% \cr
\hline
Background from misidentified muons & 0.5\% \cr 
\hline
Efficiency for intrinsic $\nue/\anue$ contamination& 80\% \cr
\hline
Signal error (systematic) & 5\% \cr
\hline
Background error (systematic) & 5\% \cr
\hline\hline
\end{tabular}
\caption{\label{tab:detector}
Detector characteristics used in the simulations. The last two rows show the reference choice of systematic errors on signal and background.
Details can be found in Section~\ref{sec:LArTPC}.}
\end{center}
\end{table}

Over the last several years, dedicated R\&D has been pursued to have a new giant next-generation and multi-purpose 
neutrino observatory based on the LArTPC technology with a total mass in the range of 
$100\,\mathrm{kt}$~\cite{Rubbia:2004tz,Rubbia:2009md,GilBotella:2004bv,Cocco:2004ac,Meregaglia:2006du,Bueno:2007um,Badertscher:2008bp}, 
devoted to particle and astroparticle physics, providing new and unique scientific opportunities in this field and likely leading to fundamental discoveries.

The LArTPC is a powerful neutrino detector for uniform and high accuracy imaging of massive active volumes~\cite{Amerio:2004ze}. 
It is based on the fact that in highly pure Argon, ionization tracks can be drifted over distances of the order of meters. 
Imaging is provided by position-segmented electrodes at the end of the drift path, continuously recording the 
signals induced. Both single and double-phase LArTPCs~\cite{Badertscher:2010zg} have been successfully operated.
Reference event timing ($T_0$) is provided by the prompt scintillation light. Application of this technology, originally developed 
at CERN, to large detectors was pioneered by the ICARUS effort which culminated in the successful
operation of the T600 on surface~\cite{Amerio:2004ze}. Underground operation was recently reported in Ref.~\cite{Rubbia:2011ft}.
Motivated by the necessity of massive detectors for neutrino physics and proton decay search, 
several designs of large LAr detectors (much more massive than ICARUS T600) have emerged in
the literature during the last decade~\cite{Cline:2001pt,Rubbia:2004tz,Rubbia:2009md,Bartoszek:2004si,Cline:2006st,Baibussinov:2007ea,Angeli:2009zza}.

In our study, the target mass of the detector is not fixed and it starts from a `pilot' of $10\,\mathrm{kt}$ and can go up to $100\,\mathrm{kt}$. We consider 
the reconstructed neutrino energy threshold to be $500\,\mathrm{MeV}$\footnote{In a LArTPC, one can actually detect energies $<100$ keV, but at these energies the triggering becomes problematic, even in an underground location. Also, at lower energies the cross-sections are more uncertain due to nuclear effects.} for all the channels. We use 80\% detection efficiency, $\epsilon$, 
for charged current (CC) electron or positron events (see Table~\ref{tab:detector}).
For $\mu^{\pm}$ events, the efficiency is 100\%. The energy resolution of the detector is expected to be very good and we assume that
the neutrino energy will be reconstructed with a Gau\ss ian energy resolution of 15\% of $\sqrt{E/\mathrm{GeV}}$.
We have considered the $\nue$ and the $\anue$ appearance channels in our study, with the background contribution 
coming from the intrinsic $\nue$/$\anue$ contamination of the beam, 
the number of muon events which will be misidentified as electron events, and NC events. 
We assume a rejection efficiency of 99.5\% for NC events\footnote{Our choice is conservative in the 
sense that the excellent spatial resolution and the capability of observing very low-energy particles might enable a LArTPC to reject 
NC events even more efficiently.} and the correlation between the visible energy of background NC events and the neutrino energy 
is implemented by migration matrices~\cite{migration}. To estimate the backgrounds coming from the intrinsic $\nue$/$\anue$ contamination, we use the same
80\% efficiency that we have used for our signal. We also assume that there is a 0.5\% chance of misidentifying muons as electron events. 
We also add the information coming from the $\numu$/$\anumu$ disappearance channels. For these channels, NC events are the main 
source of background\footnote{The $\anumu$ ($\numu$) `wrong-sign' contamination of the beam will also act as a background for the $\numu$ ($\anumu$) disappearance channel. This background is negligible compared to the number of `right-sign' signal events and moreover, the $\numu$ and $\anumu$ disappearance oscillation probabilities are similar (except for differences due to the sub-leading matter effect terms) which is why they do not cause any problem in this study.}. 
We do not consider backgrounds due to atmospheric neutrinos as the energy of the beam events is well above that of the an average atmospheric event and also, the timing information and directionality of the beam 
are sufficient to reduce these backgrounds to an insignificant level. We also include signal and background 
normalization errors (both $5\%$), uncorrelated among all oscillation channels.

\section{Event rates for the $2290\,\mathrm{km}$ and $1540\,\mathrm{km}$ baselines}
\label{sec:event}

\begin{figure}[tp]
\centering
\includegraphics[width=0.49\textwidth]{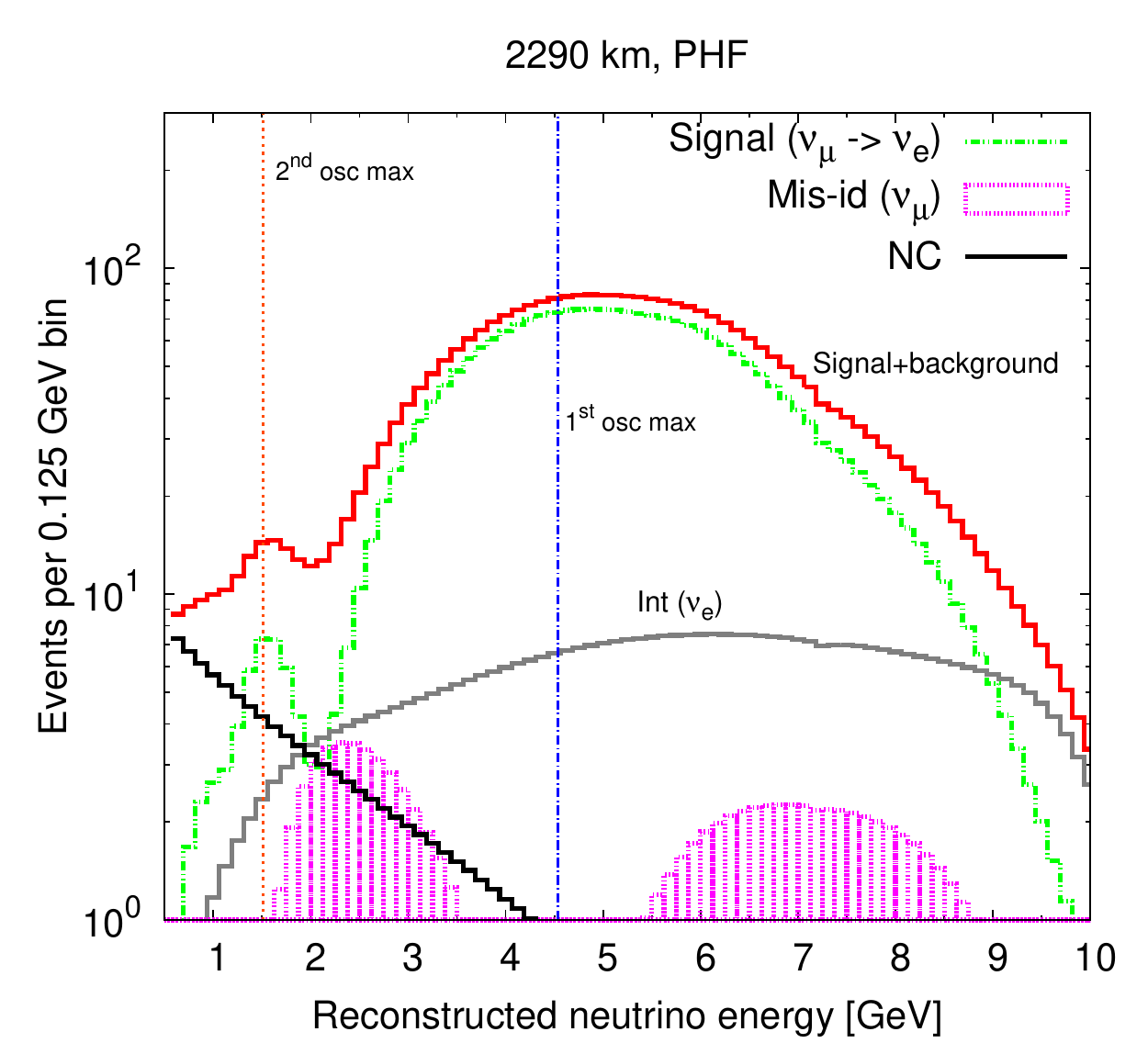}
\includegraphics[width=0.49\textwidth]{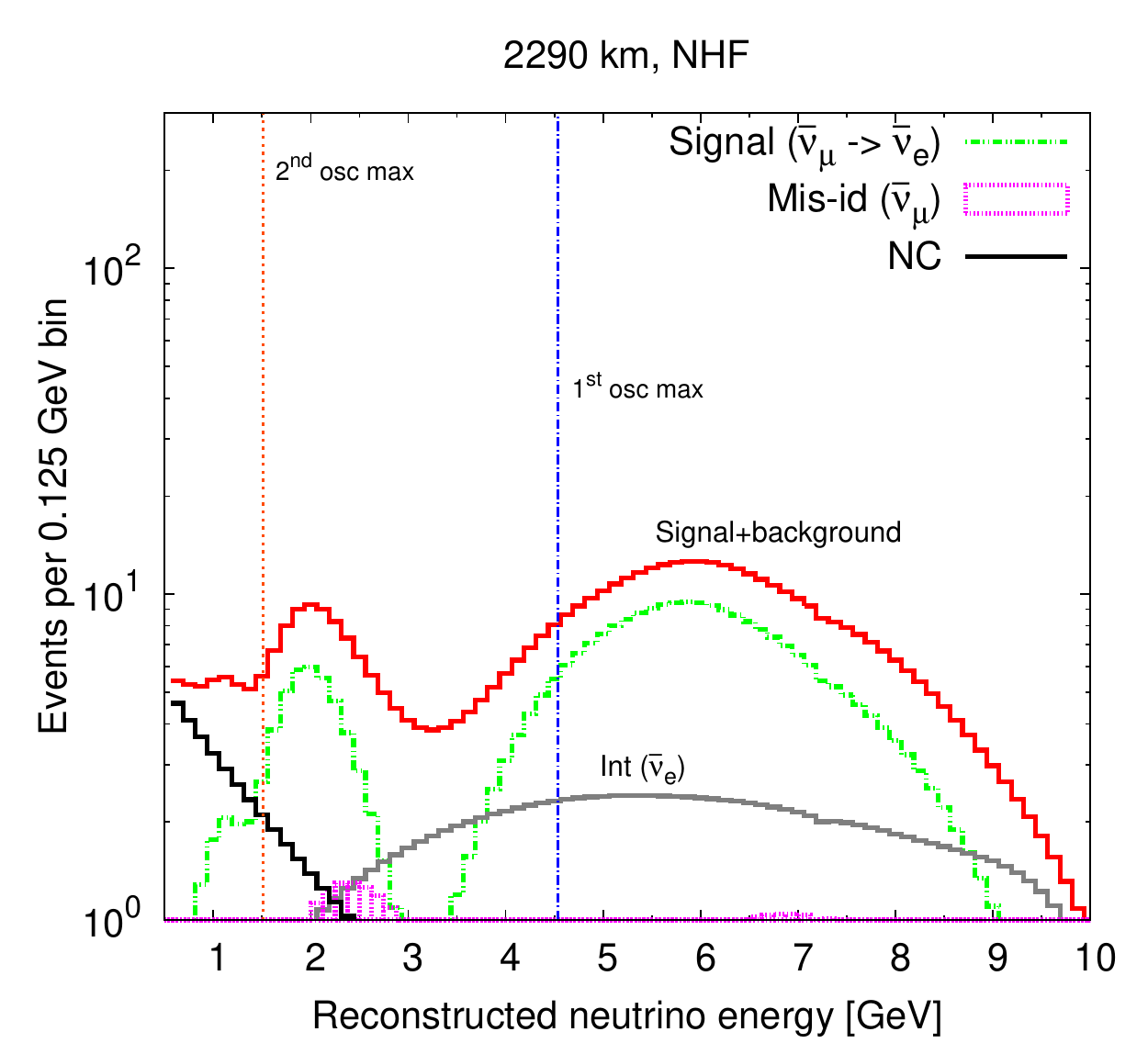}
\includegraphics[width=0.49\textwidth]{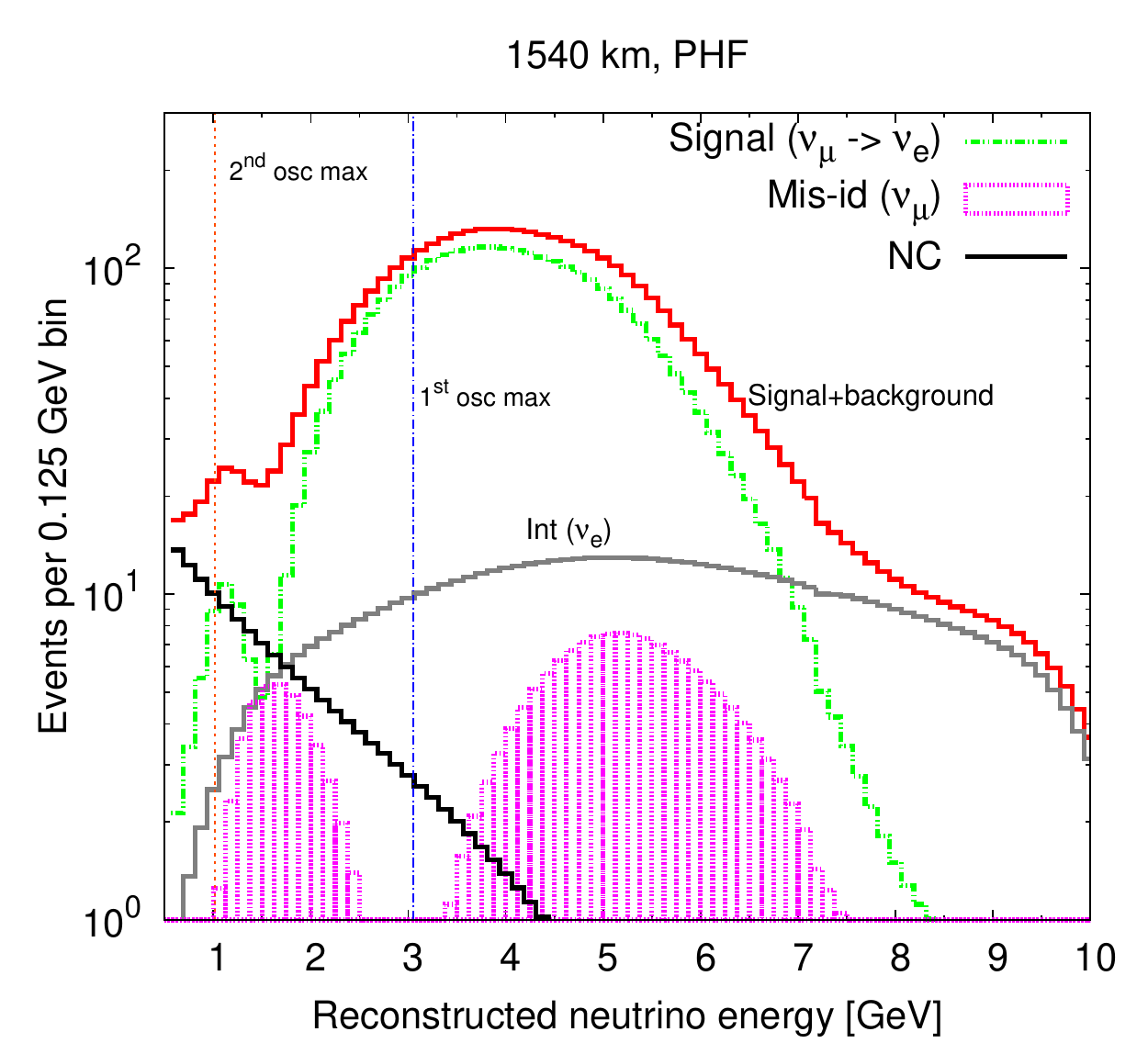}
\includegraphics[width=0.49\textwidth]{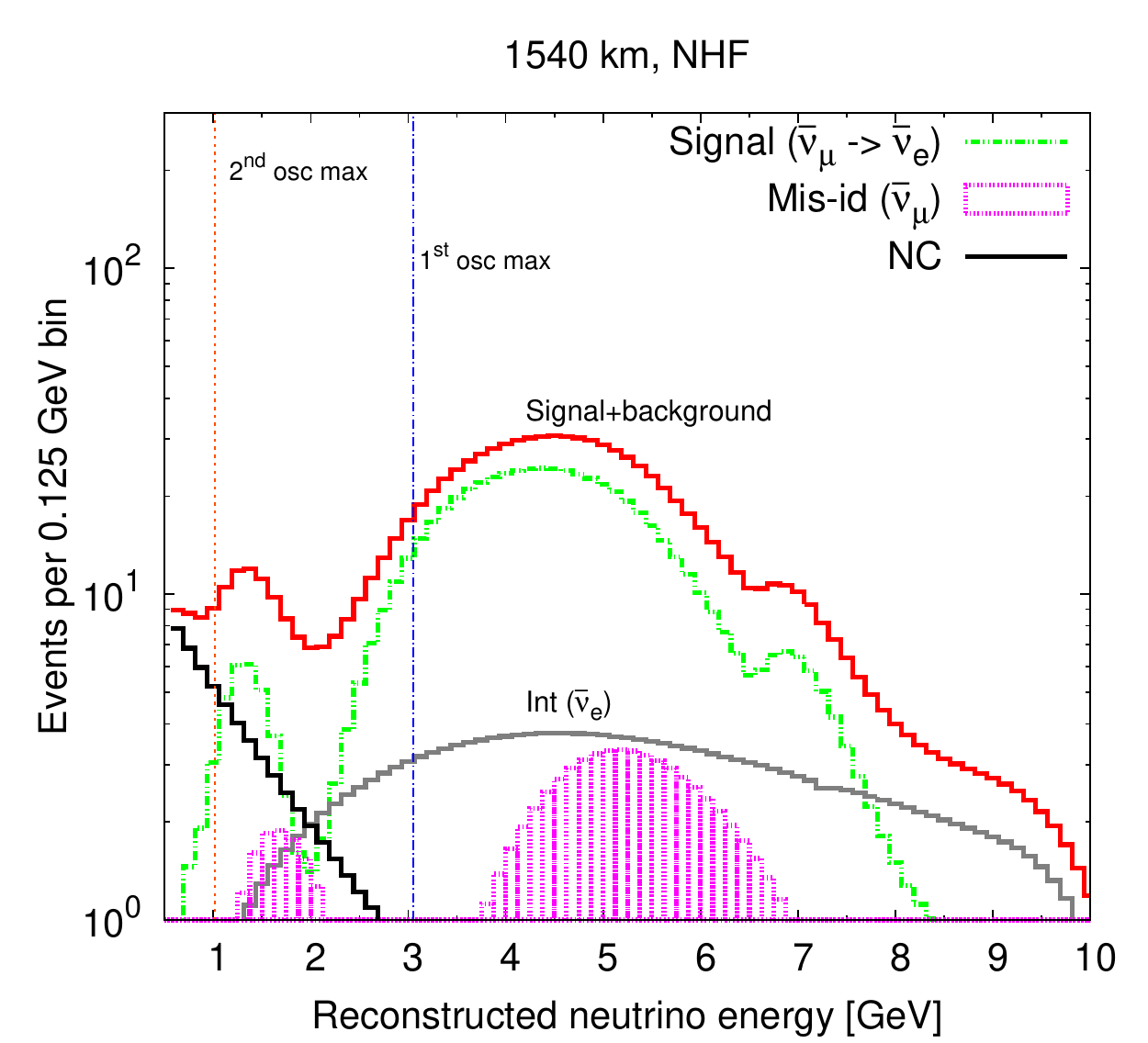}
\mycaption{\label{fig:signal-bkg} Expected signal and background event rates including the efficiency and background 
rejection capabilities in the $\nue$ (left panel)/$\anue$ (right panel) appearance channel for $\stch$ = 0.05 and $\dcp$ = 0$^{\circ}$, 
as a function of the reconstructed neutrino energy. The results are shown for an exposure of 1500 $\times$ 10$^{21}$ pot$\cdot$kt 
for each channel. The upper panels are for the CERN-Pyh{\"a}salmi baseline and the lower ones for the CERN-Slanic baseline.
A normal hierarchy has been assumed. For all other oscillation parameters we use the values given 
in Table~\ref{tab:bench}. In all the panels, the blue dot-dashed and the orange dotted vertical lines display the locations of the first and 
second oscillation maxima. The red solid histograms show the total signal$+$background rates.}
\end{figure}

In this section, we present the expected event rates in the LArTPC for the CERN-Pyh{\"a}salmi and CERN-Slanic baselines using a 
reference exposure of 1500$\times$10$^{21}$ pot$\cdot$kt (see Eq.~\eqref{eq:exposure}) for each channel.

The number of electron (positron) events\footnote{We also consider $\mu^{\pm}$ events. The number of $\mu^{\pm}$ events can be 
calculated using Eq.~\eqref{eq:events}, by making appropriate changes to the oscillation probability and cross-sections.} in the $i$-th 
energy bin in the detector is given by
\be
N_{i} = \frac{T\, n_n\, \epsilon}{4\pi L^2}~ \int_0^{E_{\rm max}}
dE \int_{E_{A_i}^{\rm min}}^{E_{A_i}^{\rm max}} dE_A \,\phi(E)
\,\sigma_\nue(E) \,R(E,E_A)\, P_{\mu e}(E) \, ,
\label{eq:events}
\ee
where $T$ is the total running time, $n_n$ is the number of target
nucleons in the detector, $\epsilon$ is the detector efficiency and
$R(E,E_A)$ is the Gau\ss ian energy resolution function of the detector.
For electron (positron) events, $\sigma_\nue$ is the neutrino (anti-neutrino) interaction cross-section. These cross-sections have
been taken from Refs.~\cite{Messier:1999kj,Paschos:2001np} and they are estimated for water and isoscalar targets. 
However, we will need to correctly evaluate the cross-sections for argon when we analyze the real experiment.
The quantities $E$ and $E_A$ are the true and reconstructed (anti-)neutrino energies respectively, and
$L$ is the baseline. 

Fig.~\ref{fig:signal-bkg} portrays the expected signal and background event spectra as a function of reconstructed neutrino energy 
including the efficiency and background rejection capabilities for the $2290\,\mathrm{km}$ (upper panels) and $1540\,\mathrm{km}$ (lower panels) 
baselines, using an exposure of 1500$\times$10$^{21}$ pot$\cdot$kt in each channel. Results are presented for $\stch$ = 0.05 and $\dcp$ = 0$^{\circ}$ 
with a NH; for all other oscillation parameters, we have used the benchmark true values as given in Table~\ref{tab:bench}. 
The left panels show the performance of the $\nue$ appearance channel and the right ones are for the $\anue$ appearance channel. 
In all the panels of Fig.~\ref{fig:signal-bkg}, one can clearly see a systematic downward bias in the reconstructed energy for NC 
background events due to the final state neutrino included via the migration matrices. The blue dot-dashed and the 
orange dotted vertical lines display the locations of the first and second oscillation maxima. The green double-dotted-dashed 
histogram shows the signal event rate and the red solid histogram presents the sum of
signal and background rate. Although we have enough statistics around the second oscillation maximum for both the baselines, its impact is 
limited due to the fact that the event samples are highly contaminated with NC and other backgrounds at lower energies.

\begin{table}[t]
\begin{center}
{\footnotesize
\begin{tabular}{|c||c|c||c|c||c|} \hline\hline
\multirow{2}{*}{Channel} & \multicolumn{2}{|c||}{{\rule[0mm]{0mm}{6mm}CERN-Pyh{\"a}salmi ($2290\,\mathrm{km}$)}}
& \multicolumn{2}{|c||}{\rule[-3mm]{0mm}{6mm}{CERN-Slanic ($1540\,\mathrm{km}$)}} \cr
\cline{2-5} 
& Signal & Background  & Signal & Background \cr
\cline{2-5}
& CC & Int+Mis-id+NC = Total  & CC & Int+Mis-id+NC = Total \cr
\hline\hline
$\numu \to \nue$ (NH) & {\bf 2364} & 419+100+103={\bf 622} & {\bf 3027} & 694+210+158={\bf 1062} \cr
\hline
$\numu \to \nue$ (IH) & {\bf 485} & 439+100+103={\bf 642} & {\bf 1110} & 710+212+158={\bf 1080} \cr
\hline
$\anumu \to \anue$ (NH) & {\bf 304} & 128+42+45={\bf 215} & {\bf 622} & 192+87+66={\bf 345} \cr
\hline
$\anumu \to \anue$ (IH) & {\bf 1049} & 122+43+45= {\bf 210} & {\bf 1127} & 188+89+66={\bf 343} \cr
\hline\hline
\end{tabular}
}
\mycaption{\label{tab:appearance}
Comparison of the total signal and background event rates in the $\nue$/$\anue$ appearance channel for CERN-Pyh{\"a}salmi and
CERN-Slanic baselines taking $\stch$ = 0.05 and $\dcp$ = 0$^{\circ}$. The results are shown for both the choices of the neutrino mass ordering
with an exposure of $1500\times10^{21}$ pot$\cdot$kt in each channel. Here signal efficiency and background rejection factors have been taken into
account from Table~\ref{tab:detector}. Here `Int' means intrinsic beam contamination, `Mis-id' means misidentified muon
events and `NC' stands for neutral current.} 
\end{center}
\end{table}

\begin{table}[t]
\begin{center}
{\footnotesize
\begin{tabular}{|c||c|c||c|c||c|} \hline\hline
\multirow{2}{*}{Channel} & \multicolumn{2}{|c||}{{\rule[0mm]{0mm}{6mm}CERN-Pyh{\"a}salmi ($2290\,\mathrm{km}$)}}
& \multicolumn{2}{|c||}{\rule[-3mm]{0mm}{6mm}{CERN-Slanic ($1540\,\mathrm{km}$)}} \cr
\cline{2-5} 
& Signal & Background  & Signal & Background \cr
\cline{2-5}
& CC & NC & CC & NC \cr
\hline\hline
$\numu \to \numu$ (NH) & 19938 & 103 & 41962 & 158 \cr
\hline
$\numu \to \numu$ (IH) & 20096 & 103 & 42486 & 158 \cr
\hline
$\anumu \to \anumu$ (NH) & 8488 & 45 & 17488 & 66 \cr
\hline
$\anumu \to \anumu$ (IH) & 8545 & 45 & 17729 & 66 \cr
\hline\hline
\end{tabular}
}
\mycaption{\label{tab:disappearance}
Comparison of the total signal and background event rates in the $\numu$/$\anumu$ disappearance channel for the CERN-Pyh{\"a}salmi and
CERN-Slanic baselines taking $\stch$ = 0.05 and $\dcp$ = 0$^{\circ}$. For all other oscillation parameters we use the values given 
in Table~\ref{tab:bench}. The results are shown for both choices of the neutrino mass hierarchy
with an exposure of $1500\times10^{21}$ pot$\cdot$kt in each channel. Signal efficiency and background rejection factors have been taken into
account from Table~\ref{tab:detector}.} 
\end{center}
\end{table}

In Table~\ref{tab:appearance}, we present a comparison between the  
total signal and background event rates in the $\nue$/$\anue$ appearance channel for the CERN-Pyh{\"a}salmi and
CERN-Slanic baselines, taking $\stch$ = 0.05 and $\dcp$ = 0$^{\circ}$. The results are shown for both choices of the 
mass hierarchy with an exposure of 1500$\times$10$^{21}$ pot$\cdot$kt in each channel, including
the signal efficiency and background rejection factors from Table~\ref{tab:detector}.
The Earth matter effects in the neutrino channel with a NH drives the number of expected events to be very large, compared to the number
expected for an IH. Similarly, in the anti-neutrino channel, matter effects in the case of an IH mean that the number of predicted
events is larger than for a NH. At the $2290\,\mathrm{km}$ baseline, the impact of matter effects is more significant compared to the
$1540\,\mathrm{km}$ baseline, and therefore the relative difference between the number of events for NH and IH is larger  
for the CERN-Pyh{\"a}salmi than the CERN-Slanic baseline. It is clear from Table~\ref{tab:appearance} that the most dominant 
contribution to the background comes from the intrinsic  $\nue$/$\anue$ beam contamination. Table~\ref{tab:disappearance}
depicts the total signal and background event rates in the $\numu$/$\anumu$ disappearance channel for the CERN-Pyh{\"a}salmi and
CERN-Slanic baselines considering an exposure of 1500$\times$10$^{21}$ pot$\cdot$kt in each channel. The information from the
 $\numu$/$\anumu$ disappearance channel is crucial to further constrain the atmospheric parameters.
For this channel, the $1540\,\mathrm{km}$ baseline has almost twice the number of events that we have in case of $2290\,\mathrm{km}$.

\begin{figure}[tp]
\includegraphics[width=0.49\textwidth]{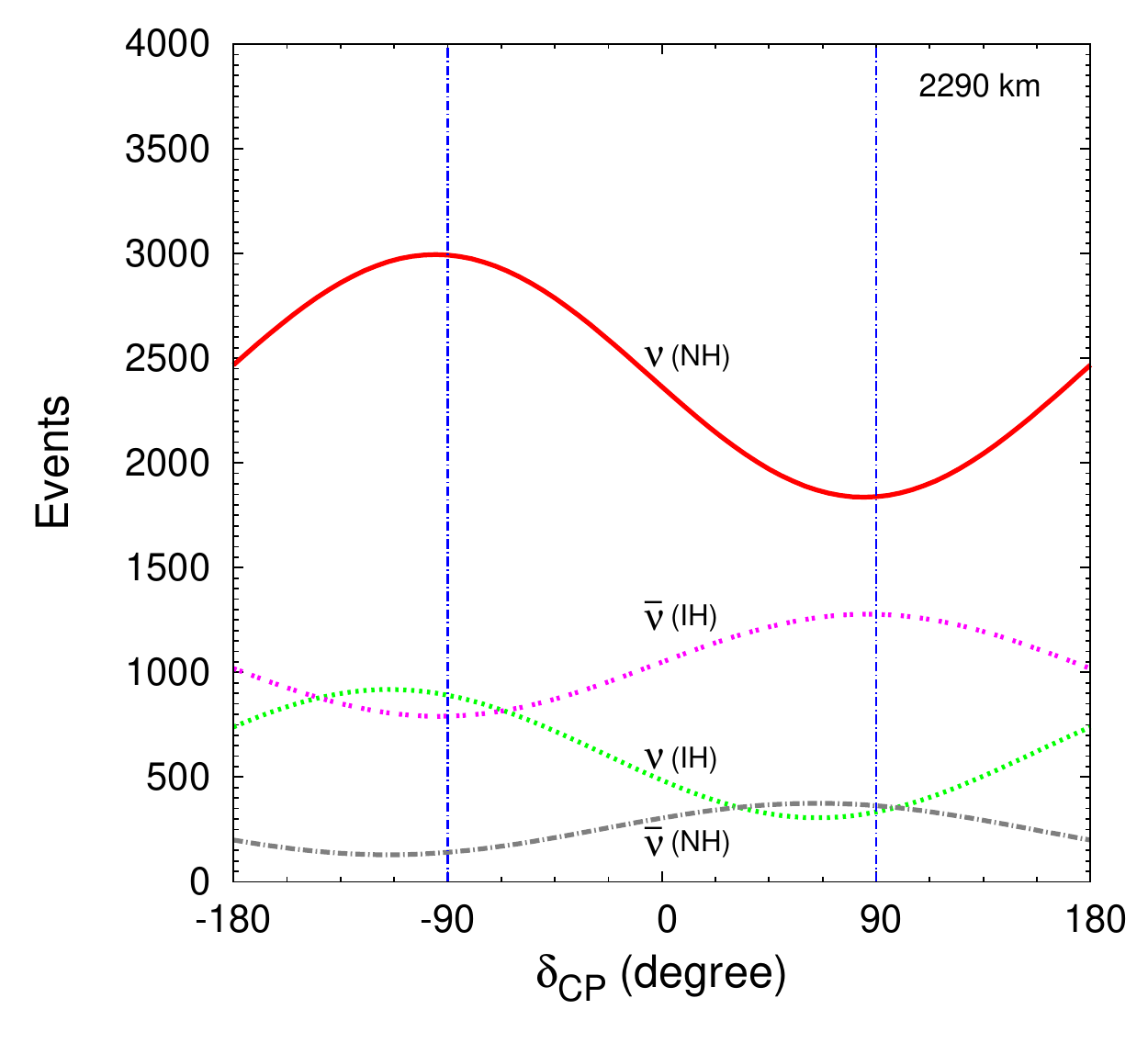}
\includegraphics[width=0.49\textwidth]{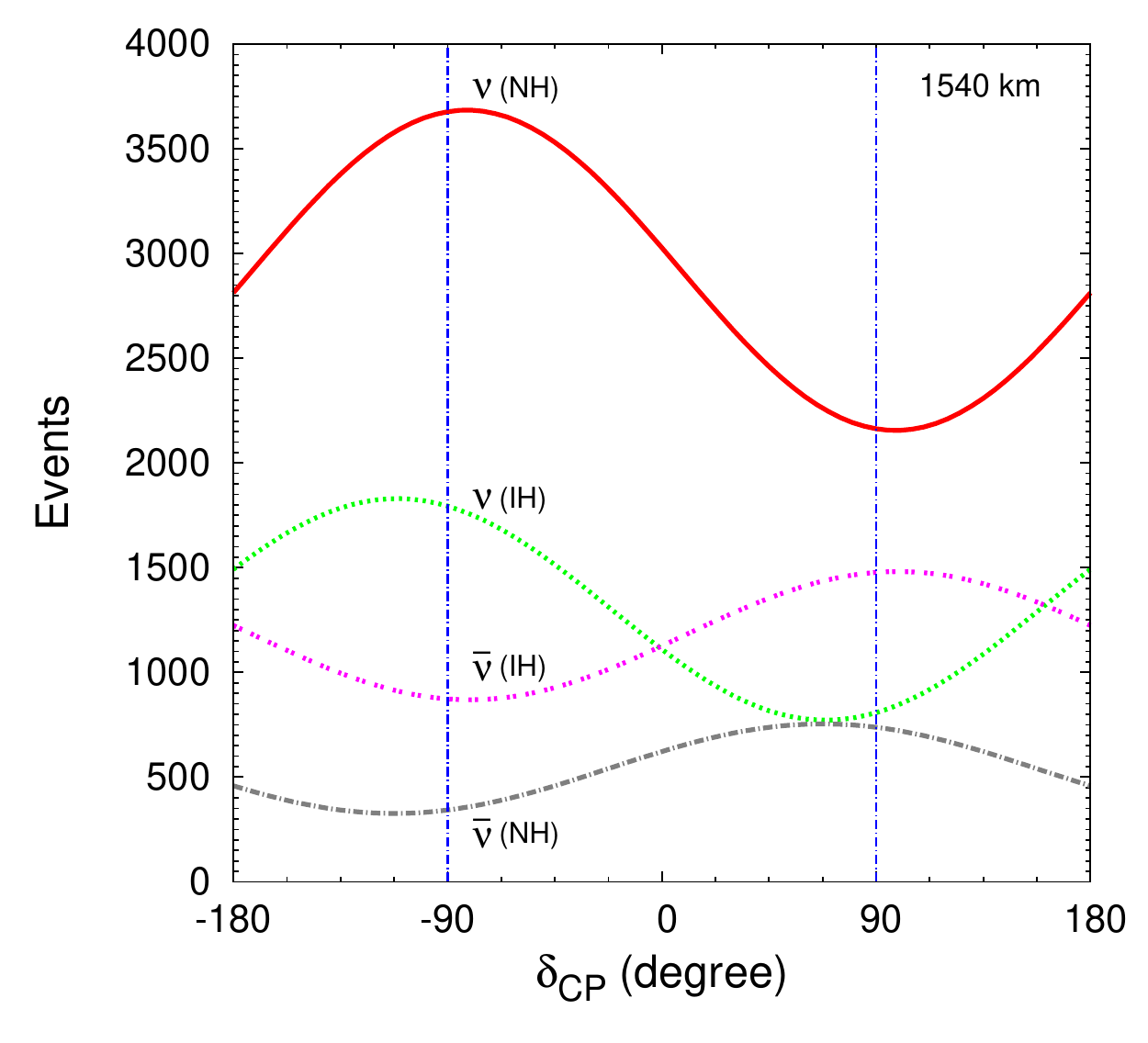}
\mycaption{\label{fig:event-cp} Event rates in the appearance channel as a function of $\dcp$. We consider $\stch$ = 0.05.
Left panel (right panel) portrays the event rates for the CERN-Pyh{\"a}salmi (CERN-Slanic) baseline.
Here, $\nu$ (NH) refers to the $\numu \to \nue$ channel with a NH. Similarly, $\bar\nu$ (NH) corresponds to the $\anumu \to \anue$ channel with a NH.
The results are shown for an exposure of 1500$\times$10$^{21}$ pot$\cdot$kt in each case.}
\end{figure}

In Table~\ref{tab:appearance}, we show the signal event rates in the $\nue$/$\anue$ appearance channel for a specific value of
$\dcp$ which we have chosen to be one of the CP conserving cases. It is interesting to see how the signal event rate varies with the choice of
$\dcp$, which is what we show in Fig.~\ref{fig:event-cp}. This shows the $\dcp$ dependence of the event rates for the CERN-Pyh{\"a}salmi (left panel) and 
CERN-Slanic (right panel) baselines. In each panel, the results are shown for both the $\nue$ and $\anue$ appearance 
channels considering both choices of the mass hierarchy. Here, we consider $\stch$ = 0.05 and an exposure of 1500$\times$10$^{21}$ 
pot$\cdot$kt in each case. The number of events varies in a wide range with the choice of $\dcp$ for both the baselines.
We can see that the maximum (minimum) number of events for neutrinos is obtained around -90$^\circ$ (90$^\circ$) for both
NH and IH. For anti-neutrinos, the same is true with $\dcp \rightarrow -\dcp$.
We can analyze this behavior with the help of Eq.~\eqref{eq:pmue}. We do not have resonant matter effects for either of the baselines and 
hence $\hat{A} < 1$ in Eq.~\eqref{eq:matt}. It follows that the $\dcp$ dependence of the probabilities can be written as 
$P_{\mu e} \simeq C_0 + C_1 - C_- \sin \dcp + C_+ \cos \dcp$ and $P_{\bar{\mu}\bar{e}} \simeq C_0 + C_1 + C_- \sin \dcp + C_+ \cos \dcp$, 
where $C_0, C_1$ and, $C_\pm$ are independent of $\dcp$, whence the symmetry is manifest. For most of the energies, $C_-$ is positive assuming a NH.  
Now for $\dcp=-90^\circ$ (90$^\circ$) the term $C_- \sin \dcp$ gives a positive (negative) contribution towards the probability for neutrinos. 
The opposite is true for anti-neutrinos. A similar argument is valid for the case of an IH.  

\section{Simulation method}
\label{sec:method}

\begin{table}[t]
\begin{center}
\begin{tabular}{||c||c||} \hline\hline
\multicolumn{1}{||c||}{{\rule[0mm]{0mm}{6mm}{Benchmark Values}}}
& \multicolumn{1}{|c||}{\rule[-3mm]{0mm}{6mm}{$1\sigma$ estimated error}} 
\cr
\hline \hline
$\Delta m^2_{31}{\rm (true)} = 2.45 \times 10^{-3} \ {\rm eV}^2$ (NH) & $\sigma(\Delta m^2_{31})= 5\%$ \cr
\hline
$\Delta m^2_{31}{\rm (true)} = -2.34 \times 10^{-3} \ {\rm eV}^2$ (IH) & $\sigma(\Delta m^2_{31})= 5\%$ \cr
\hline
$\theta_{23}{\rm (true)} = 45^{\circ}$ & $\sigma(\theta_{23})=10\%$ \cr 
\hline
$\Delta m^2_{21}{\rm (true)} = 7.59 \times 10^{-5} \ {\rm eV}^2$ & $\sigma(\Delta m^2_{21})=3\%$ \cr 
\hline
$\theta_{12}{\rm (true)} = 33.96^{\circ}$ & $\sigma(\theta_{12})=3\%$ \cr 
\hline
$\rho{\rm (true)} = 1$ & $\sigma(\rho)=2\%$ \cr
\hline \hline
\end{tabular}
\mycaption{\label{tab:bench}
Chosen benchmark values of oscillation parameters and their $1\,\sigma$ estimated errors.
In the last row, $\rho$ is the Earth matter density, relative to the value given by the Preliminary Reference Earth Model (PREM).}
\end{center}
\end{table}

For all the numerical results presented in this study, we have used the GLoBES software~\cite{Huber:2004ka,Huber:2007ji}. 
Unless stated otherwise, we have generated our simulated data for the benchmark values given in the first column of 
Table~\ref{tab:bench}. These values correspond to the current best-fit values of the oscillation parameters,
in light of the current neutrino data~\cite{Schwetz:2011qt}. In all fits, these parameters are marginalized over, with the
$1\,\sigma$ intervals given in the second column of Table~\ref{tab:bench} included as priors\footnote{The setup we are considering can improve upon the current measurements of the
atmospheric parameters. However, it is useful to add the current informatio to speed 
up the marginalization.} (see Eq.~\eqref{eq:prior}). We also include a 2\% uncertainty on the matter density ($\rho$)~\cite{Geller:2001ix,Ohlsson:2003ip,Kozlovskaya:2003kk}. 

For our statistical analysis we use the techniques outlined in Refs.~\cite{Huber:2002mx,Fogli:2002au} and employ the following $\chi^2$ function:
\be
\chi^2_{\rm total} &=&
\chi^2_{\numu \rightarrow \nue} + \chi^2_{\anumu \rightarrow \anue} + \chi^2_{\numu \rightarrow \numu} + \chi^2_{\anumu \rightarrow \anumu} + \chi^2_{\rm prior}.
\label{eq:totchisq}
\ee
The $\chi^2$ for the $\numu \rightarrow \nue$ appearance channel is given by
\be
\chi^2_{\numu \rightarrow \nue} = min_{\xi_s, \xi_b}\left[2\sum^{n}_{i=1}
(\tilde{y}_{i}-x_{i} - x_{i} \ln \frac{\tilde{y}_{i}}{x_{i}}) +
\xi_s^2 + \xi_b^2\right ]~,
\label{eq:chipull}
\ee
where $n$ is the total number of bins and
\be
\tilde{y}_{i}(\{\omega\},\{\xi_s, \xi_b\}) = N^{th}_i(\{\omega\}) \left[
1+ \pi^s \xi_s \right] +
N^{b}_i(\{\omega\}) \left[1+ \pi^b \xi_b \right]~.
\label{eq:rth}
\ee
In the above expression, $N^{th}_i(\{\omega\})$ is the predicted number of CC signal events (calculated using Eq.~\eqref{eq:events}) 
in the $i$-th energy bin for a set of oscillation parameters $\omega$ and $N^{b}_i(\{\omega\})$ are the total number of background 
events\footnote{We consider both CC and NC background events; NC events do not depend on the oscillation parameters.} 
in bin $i$. The quantities $\pi^s$ and $\pi^b$ in Eq.~\eqref{eq:rth} are the systematic errors on the
signals and backgrounds respectively. We consider $\pi^s = 5\%$ and $\pi^b = 5\%$ as mentioned in 
Table~\ref{tab:detector}. The quantities $\xi_s$ and $\xi_b$ are the pulls due to the systematic error on the signal and
background respectively. The data in Eq.~\eqref{eq:chipull} enters through the variable $x_i=N_i^{ex}+N_i^b$, 
where $ N_i^{ex}$ is the number of observed CC signal events in the detector and $N_i^b$ is the background, 
as mentioned earlier. We simulate the CC signal event spectrum using Eq.~\eqref{eq:events} for our true values of the oscillation 
parameters given in the first column of Table~\ref{tab:bench}. We have chosen to study a range of values of   
$\theta_{13}$ such that $0.01 < \stch < 0.1$, consistent with the $2\,\sigma$ range as given in Ref.~\cite{Schwetz:2011zk}. 
We consider the full range of values of $\dcp$ and study both the mass hierarchies. In a similar fashion, we 
calculate the contributions towards $\chi^2_{total}$ coming from the other oscillation channels (for both neutrino and anti-neutrino modes).
In our $\chi^2$ fit we marginalize over {\it all} oscillation parameters, as well as the mass hierarchy, by allowing these parameters to vary in the fit and picking the smallest value of the $\chi^2$ function.
However, we impose a prior, or external constraint, on the atmospheric and solar parameters and on the matter density, 
through a $\chi^2_{prior}$, given by
\be
\chi^2_{prior} &=& \left (\frac{|\Delta m^2_{31}|-
    |\mat|}{\sigma(|\Delta m^2_{31}|)} \right )^2 +
\left (\frac{\tmt-\tmt{\mbox {(true)}}}{\sigma(\tmt)} \right )^2\nonumber \\
&+& \left (\frac{\Delta m^2_{21}- \mst}{\sigma(\Delta m^2_{21})}
\right )^2 + \left (\frac{\tem-\tem{\mbox {(true)}}}{\sigma(\tem)} \right )^2\nonumber \\
&+&
\left (\frac{\rho-1}{\sigma(\rho)} \right )^2 ~,
\label{eq:prior}
\ee
where $\rho$ is the matter density relative to the PREM density, and the $1\sigma$ errors are those given in the second column of Table~\ref{tab:bench}.
No priors are included for $\tet$ and $\dcp$ which means that they are completely free in the marginalization.
However, it is expected that the currently running or upcoming accelerator and reactor experiments will not only strengthen the emerging hints of non-zero
$\tet$ but also provide an allowed range of $\stch$ at high significance, independent of the value of $\dcp$. 
This information will be very useful when incorporated in the form of a prior on $\tet$, and will certainly significantly improve 
the results presented in this study. 

\section{Results}
\label{sec:results}

In this section we explore the capability of the facility described in Section~\ref{sec:setup}, with both the $2290\,\mathrm{km}$ and $1540\,\mathrm{km}$ baselines, 
to make a measurement of the true neutrino mass hierarchy (MH) and the CPV phase, $\dcpt$. For long-baseline experiments, the 
measurement of the mass hierarchy is easier than a measurement of $\dcpt$ because matter effects enhance the separation between the oscillation 
spectra, and therefore the event rates, of a NH and an IH. Additionally, this measurement is one that is `discrete' as we only need to differentiate between two possibilities. 
A `discovery' of the mass hierarchy is defined as the ability to exclude any degenerate solution for the wrong (fit) hierarchy 
at a given confidence level; we consider the $3\,\sigma$ and $5\,\sigma$ levels. 
A `discovery' of CPV, if it exists, means being able to exclude the CP-conserving values 
of $0^{\circ},\,180^{\circ}$ at a given confidence level (again, we consider the $3\,\sigma$ and $5\,\sigma$ confidence levels). 
Clearly, this measurement is increasingly difficult the closer that $\dcpt$ is to $0^{\circ},\,180^{\circ}$. Therefore, whilst it is possible to 
discover the mass hierarchy for \emph{all} possible values of $\dcpt$, the same is not true for CPV. As such, we focus firstly on the measurement of the 
mass hierarchy, for which we will show that a result can be guaranteed for suitably large $\theta_{13}$, irrespective of the true value of $\dcpt$, with only a modest exposure. We will then discuss the CPV discovery potential of the facility, showing that whilst a larger exposure is needed in order to obtain a reasonable sensitivity than for a measurement of the mass hierarchy, it is still feasible to cover a significant portion of the parameter space with a moderate exposure.

We remind the reader that when we refer to `PHF' (positive horn focusing) and `NHF' (negative horn focusing) in this section, the former refers to a beam consisting predominantly of $\nu_{\mu}$'s whereas the latter refers to a beam consisting mainly of $\bar{\nu}_{\mu}$'s, as shown in Fig.~\ref{fig:flux}.

\subsection{Measurement of the mass hierarchy} 
\label{sec:mh}

\begin{figure}[t]
\centering
\includegraphics[width=0.5\textwidth]{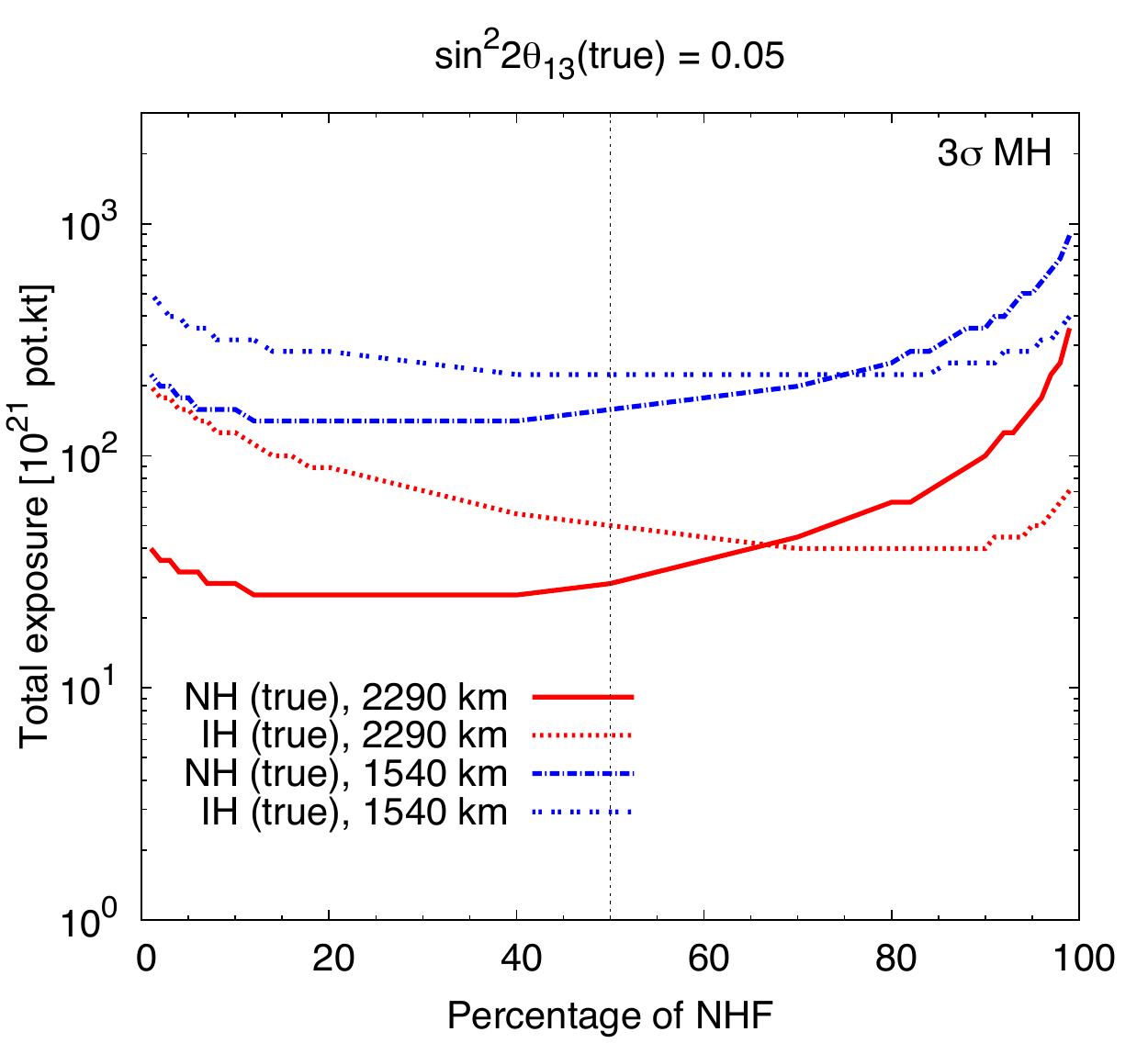}
\mycaption{\label{fig:hier-ratio}Total exposure needed to achieve mass hierarchy discovery at $3\,\sigma$ C.L. for 
$100\%$ values of $\dcpt$ as a function of the percentage of NHF in the total exposure, for $\stcht=0.05$. The results 
are qualitatively similar for other values of $\stcht$ and confidence levels.}
\end{figure}

\begin{figure}[t]
\includegraphics[width=5.75cm, height=5.75cm]{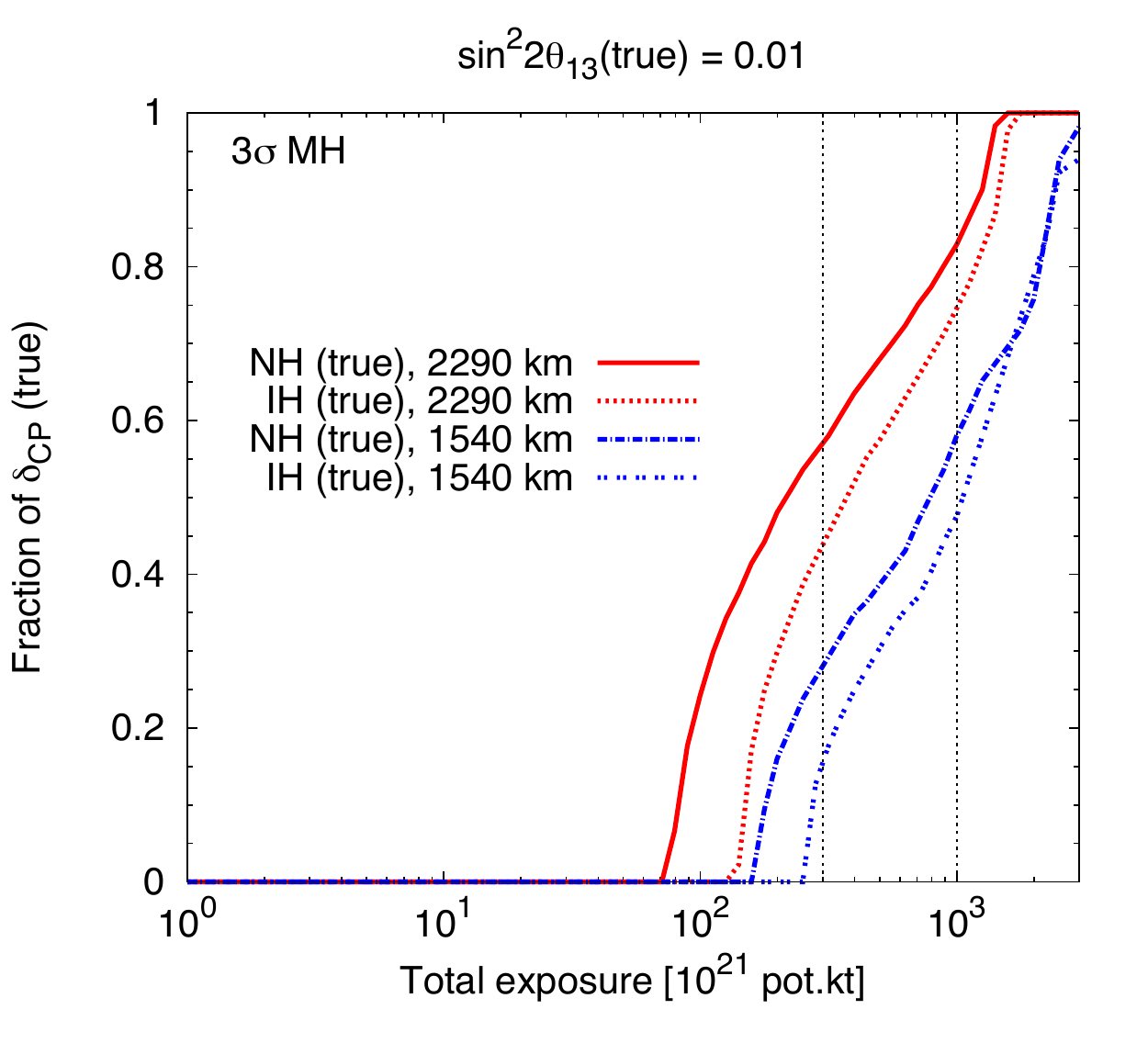}
\includegraphics[width=5.75cm, height=5.75cm]{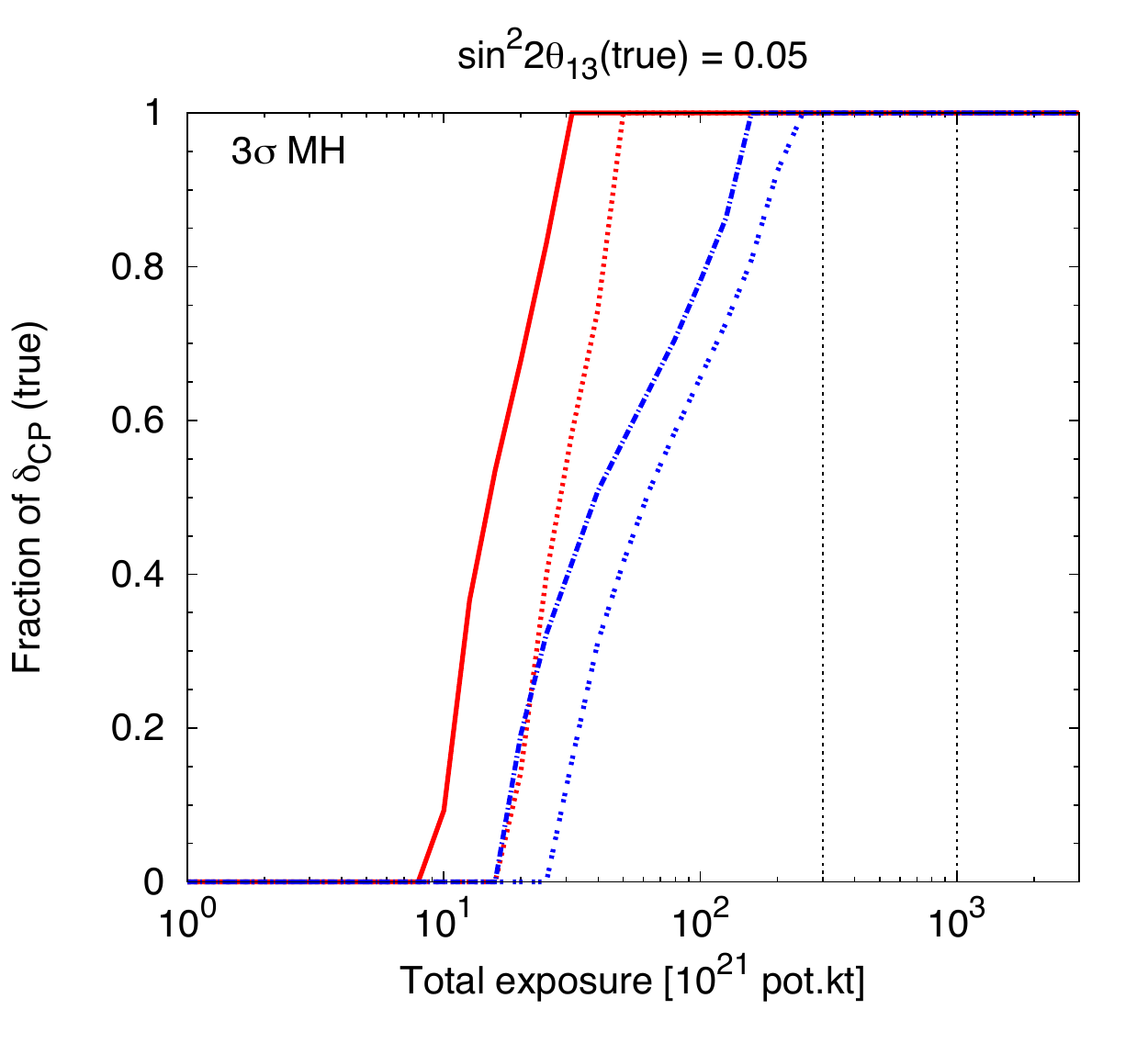}
\includegraphics[width=5.75cm, height=5.75cm]{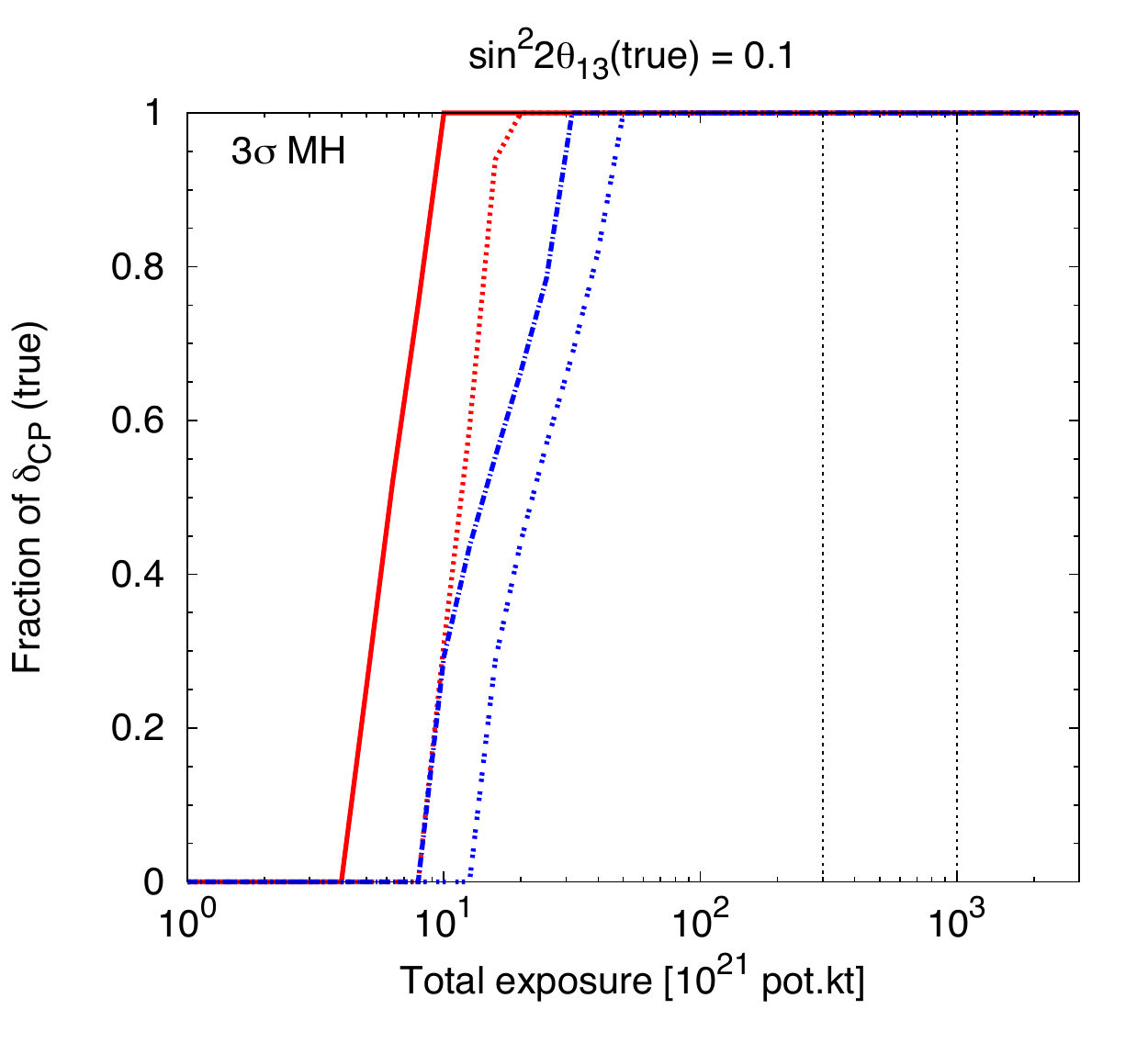}
\includegraphics[width=5.75cm, height=5.75cm]{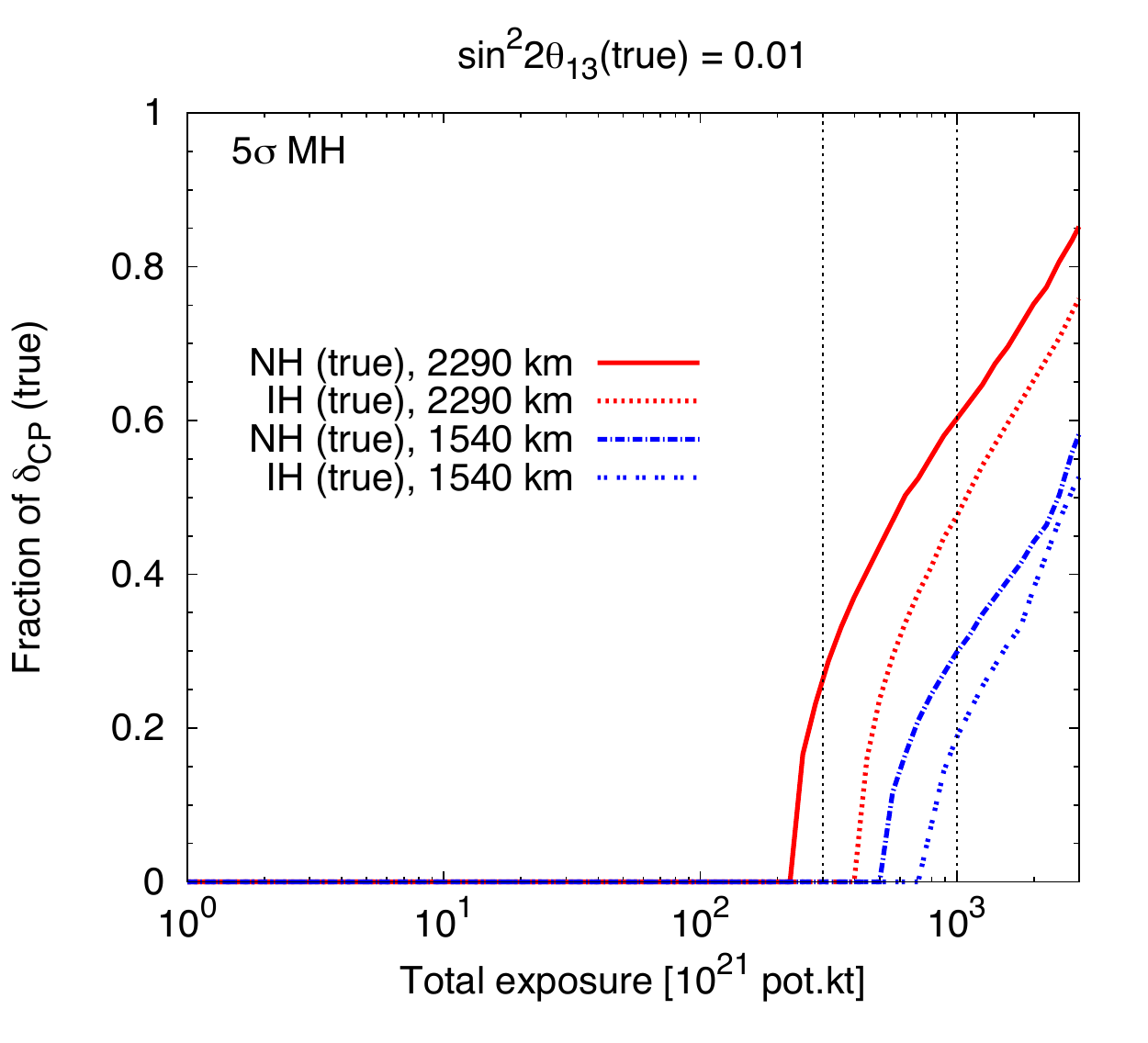}
\includegraphics[width=5.75cm, height=5.75cm]{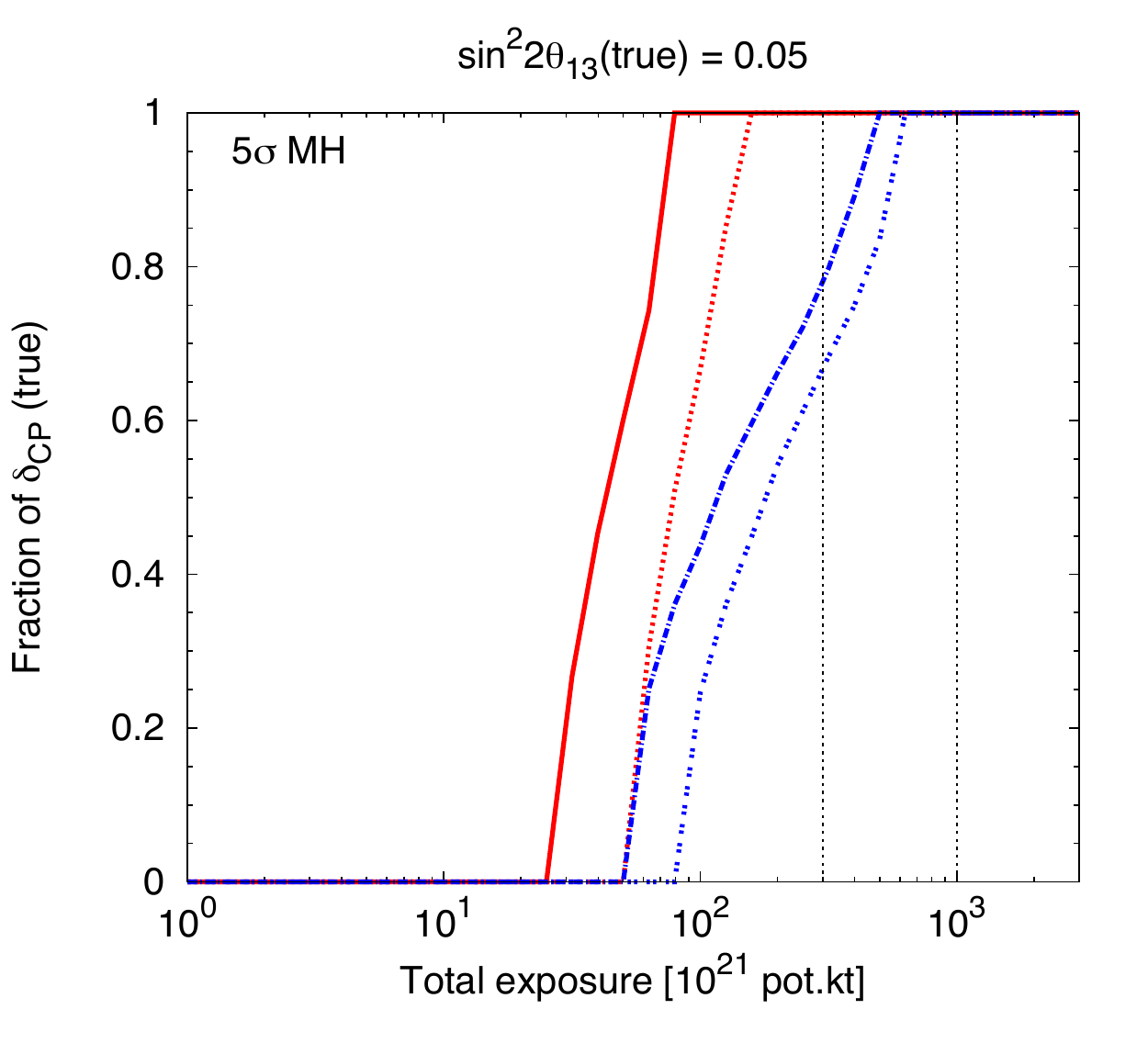}
\includegraphics[width=5.75cm, height=5.75cm]{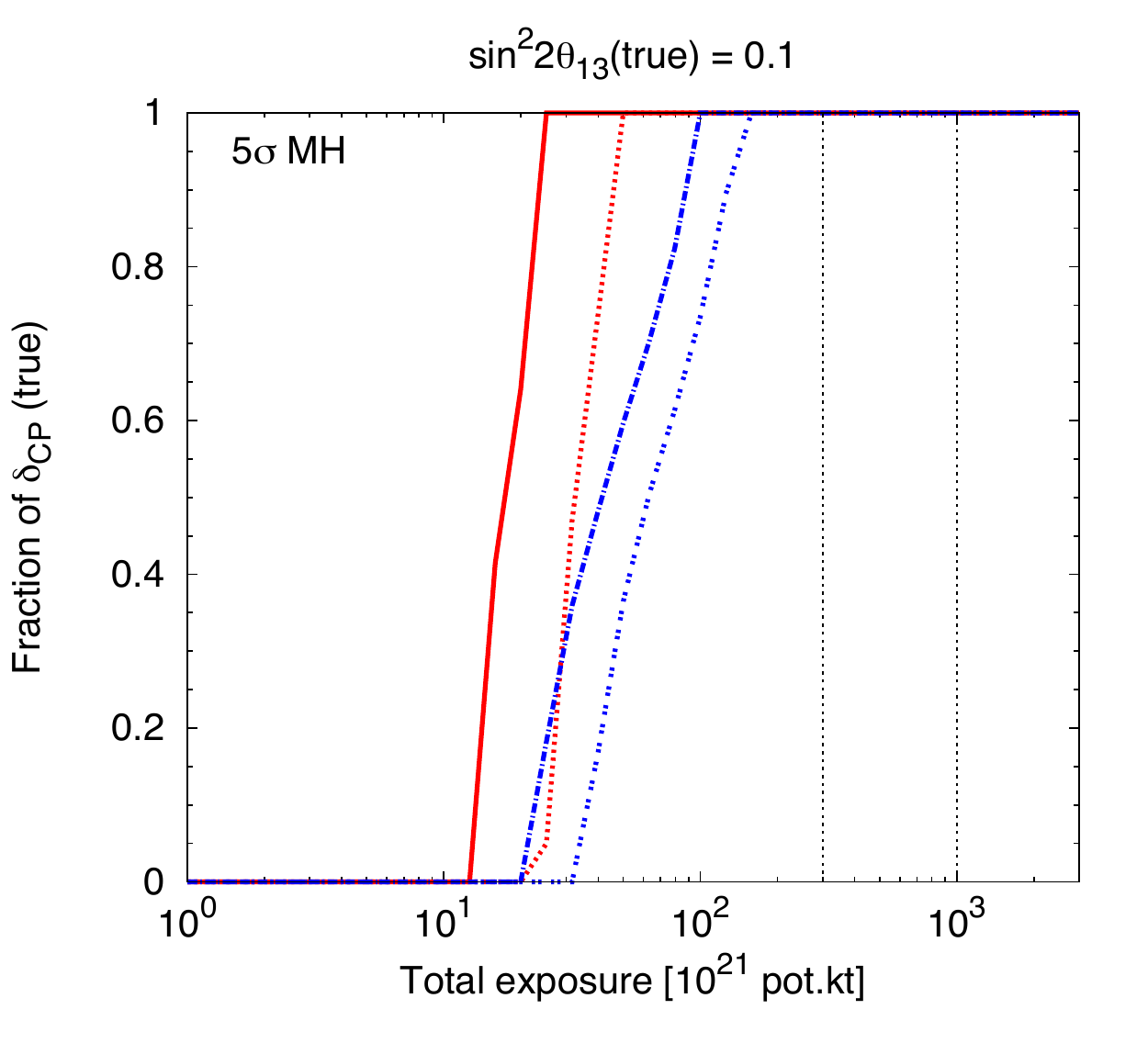}
\mycaption{\label{fig:hier-expo} CP fractions for which a mass hierarchy discovery at $3\,\sigma$ (upper panels) and $5\,\sigma$ (lower panels)
confidence levels is possible as a function of total exposure, with equal sharing of PHF and NHF. Results are depicted for three different
choices of $\stcht$: 0.01 (left panel), 0.05 (middle panel) and 0.1 (right panel).}
\end{figure}

We begin by referring back to Fig.~\ref{fig:event-cp} where we showed the event rates in the appearance channel as a function of $\dcpt$, for both a NH and an IH. As an initial simplistic analysis, we can firstly look at the relative \emph{difference} between the number of events for a NH and an IH, for a given value of $\dcpt$. This gives a rough idea as to how easy it is to distinguish between the mass hierarchies, for each of the baselines. We can see that the difference is always greater for the $2290\,\mathrm{km}$ baseline than for the $1540\,\mathrm{km}$ baseline, for all values of $\dcpt$ and both $\nu$ and $\bar{\nu}$, which is what is expected due to the stronger matter effects at the longer baseline. We also see that the distinction between the mass hierarchies is greatest for $\nu$'s around $\dcpt\sim-90^{\circ}$ whereas the minimum occurs around $\dcpt\sim+90^{\circ}$. The opposite is true for $\bar{\nu}$'s, which means that $\nu$'s and $\bar{\nu}$'s have sensitivity to complementary parts of the $\dcpt$ parameter space. Therefore, in order to minimize the exposure needed to obtain coverage for all values of $\dcpt$, it is advantageous to combine both PHF and NHF.

The question is exactly what combination of PHF:NHF is optimal, in the sense of both the physics potential and experimental feasibility. Experimentally, PHF is preferable to NHF for several reasons. Predominantly, the $\bar{\nu}$ charged-current cross-section is lower than for $\nu$ whereas the associated background is higher, and the systematic errors for $\bar{\nu}$ are currently greater than for $\nu$. Additionally, the production yield for $\pi^{-}$, which decays into $\bar{\nu}_{\mu}$, is slightly lower than for $\pi^{+}$'s. Therefore we want to know if, firstly, data from NHF is actually necessary, and secondly, if it is, the \emph{minimum} amount of NHF required for optimal performance. In Fig.~\ref{fig:hier-ratio} we show the \emph{total} exposure (PHF and NHF combined) needed in order to make a $3\,\sigma$ determination of the mass hierarchy for \emph{all} values of $\dcpt$, as a function of the percentage of the total exposure for which the experiment runs in NHF. The results are shown for both the baselines we are considering, and for both NH (true) and IH (true). The point at which the lines for NH (true) and IH (true) cross, for each baseline, shows the amount of NHF running needed such that the same total exposure is adequate for both of the mass hierarchies. Assuming that we have no prior hints as to the nature of the true hierarchy before the experiment starts running, then at all other points, the total exposure which is needed must be taken to be the higher one. Therefore, without prior knowledge about the mass hierarchy, the crossing point is the optimum.

We find that it is possible to identify the mass hierarchy for all values of $\dcpt$ with only PHF or only NHF, which indicates that there are no intrinsic degeneracies which cannot be resolved simply by accumulating more data. Approximately an order of magnitude more running is needed for NHF alone than for PHF alone, in the case of a NH (true), and vice versa for an IH (true); a NH favors more data from $\nu$'s whereas an IH prefers data from $\bar{\nu}$'s. This is a consequence of the fact that $\nu$ oscillations are enhanced by matter in the case of a NH whereas $\bar{\nu}$ oscillations are enhanced in the case of an IH.

For $2290\,\mathrm{km}$, the optimum lies at a PHF:NHF ratio of around $30:70$, whereas the shorter $1540\,\mathrm{km}$ baseline prefers slightly more NHF and requires a higher total exposure. This can be understood if we refer back to Fig.~\ref{fig:event-cp} which shows that the relative difference between the event rates for a NH and an IH is always larger for $2290\,\mathrm{km}$ than $1540\,\mathrm{km}$, and hence a lower exposure is required for the longer baseline. Additionally, as mentioned earlier, we see how $\nu$'s and $\bar{\nu}$'s have sensitivity to complementary parts of the $\dcpt$ space. Since we require a minimum threshold number of events to perform a $3\,\sigma$ determination of the mass hierarchy, for each value of $\dcpt$, this means that the $1540\,\mathrm{km}$ baseline requires more $\bar{\nu}$ data than $2290\,\mathrm{km}$ in order to reach the threshold number for the values of $\dcpt$ in which the $\nu$ data is weakest. 

Bearing in mind that $50\%$ NHF is the maximum that is experimentally favorable, and that the difference in exposure required between the exact optimum and the $50:50$ mixture is small, we set a $50:50$ ratio of PHF and NHF (shown by the dotted vertical black line) to be our reference configuration for mass hierarchy discovery. In Fig.~\ref{fig:hier-ratio} we show only the results for $\stcht=0.05$ at $3\,\sigma$ confidence but we have verified that the quantitative optimization is practically independent of the value of $\theta_{13}$(true) and the statistical confidence level.

In Fig.~\ref{fig:hier-expo} we show how the sensitivity to the mass hierarchy varies as a function of total exposure, when we use our reference configuration with equal PHF and NHF. We show the results for both baselines and mass hierarchies, and three different values of $\stcht$ - 0.01 (left panels), 0.05 (middle panels) and 0.1 (right panels). The upper row shows the results at $3\,\sigma$ C.L. and the bottom row the results for $5\,\sigma$ C.L.. The dotted vertical black lines mark exposures of $300\times10^{21}$ pot$\cdot$kt and $1000\times10^{21}$ pot$\cdot$kt, chosen for reasons that will be explained in the following subsection. The right-most edge of the graph lies at our maximum exposure of $3000\times10^{21}$ pot$\cdot$kt. For all values of $\stcht$ there is slightly better sensitivity to a NH than an IH (see Fig.~\ref{fig:hier-ratio} where the NH (true) line is at a lower exposure than the IH (true) line for $50\%$ NHF). For the longer baseline, sensitivity begins at a lower exposure and the increase in sensitivity as a function of exposure is more rapid than for the $1540\,\mathrm{km}$ baseline, although the differences are not large. Even if $\stcht\sim0.01$, both baselines can obtain $3\,\sigma$ sensitivity for the complete parameter space. The $1540\,\mathrm{km}$ baseline is capable of $100\%$ coverage at $5\sigma$ confidence for values of $\theta_{13}$ 
down to $\stcht\sim0.025$ (not shown); for the $2290\,\mathrm{km}$ baseline it is nearly possible at $\stcht\sim0.01$.

The capability of this setup then, for the large values of $\theta_{13}$ under consideration, is remarkable: if we consider $\stcht=0.1$ and the $2290\,\mathrm{km}$ baseline then complete sensitivity at $3\,\sigma$ can be gained with an exposure of only $20\times10^{21}$ pot$\cdot$kt. This corresponds to, for example, a 10 kt detector with a beam running for roughly one year. We obtain complete $5\,\sigma$ sensitivity with roughly twice this exposure. For $\stcht=0.025$, the $2290\,\mathrm{km}$ baseline is capable of achieving $100\%$ sensitivity at $3\,\sigma$ confidence after roughly $200\times10^{21}$ pot$\cdot$kt (20 kt detector running for 5 years, for example).

\subsection{CP violation discovery potential} 
\label{sec:cpv}

\begin{figure}[tp]
\includegraphics[width=5.75cm, height=5.75cm]{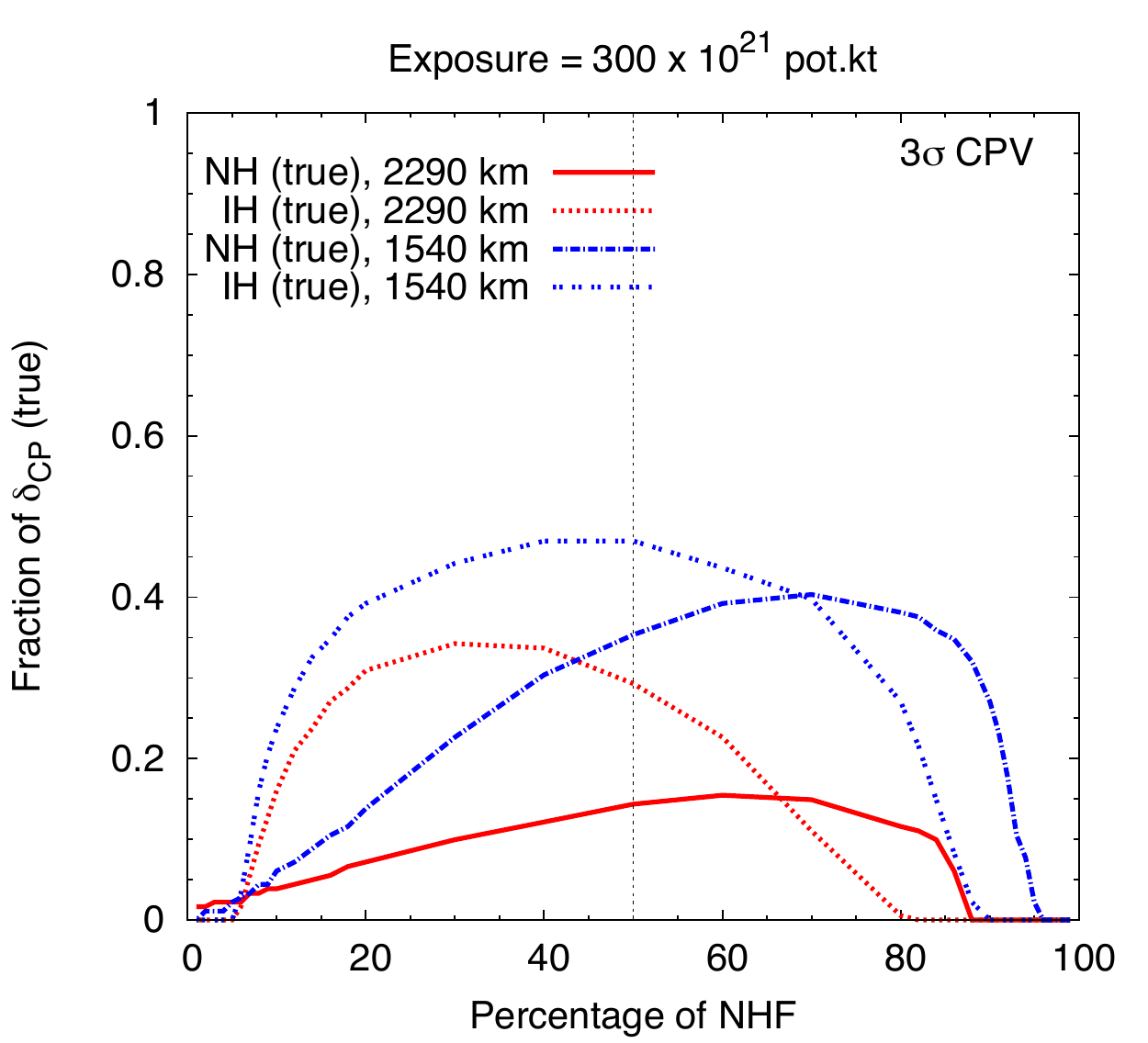}
\includegraphics[width=5.75cm, height=5.75cm]{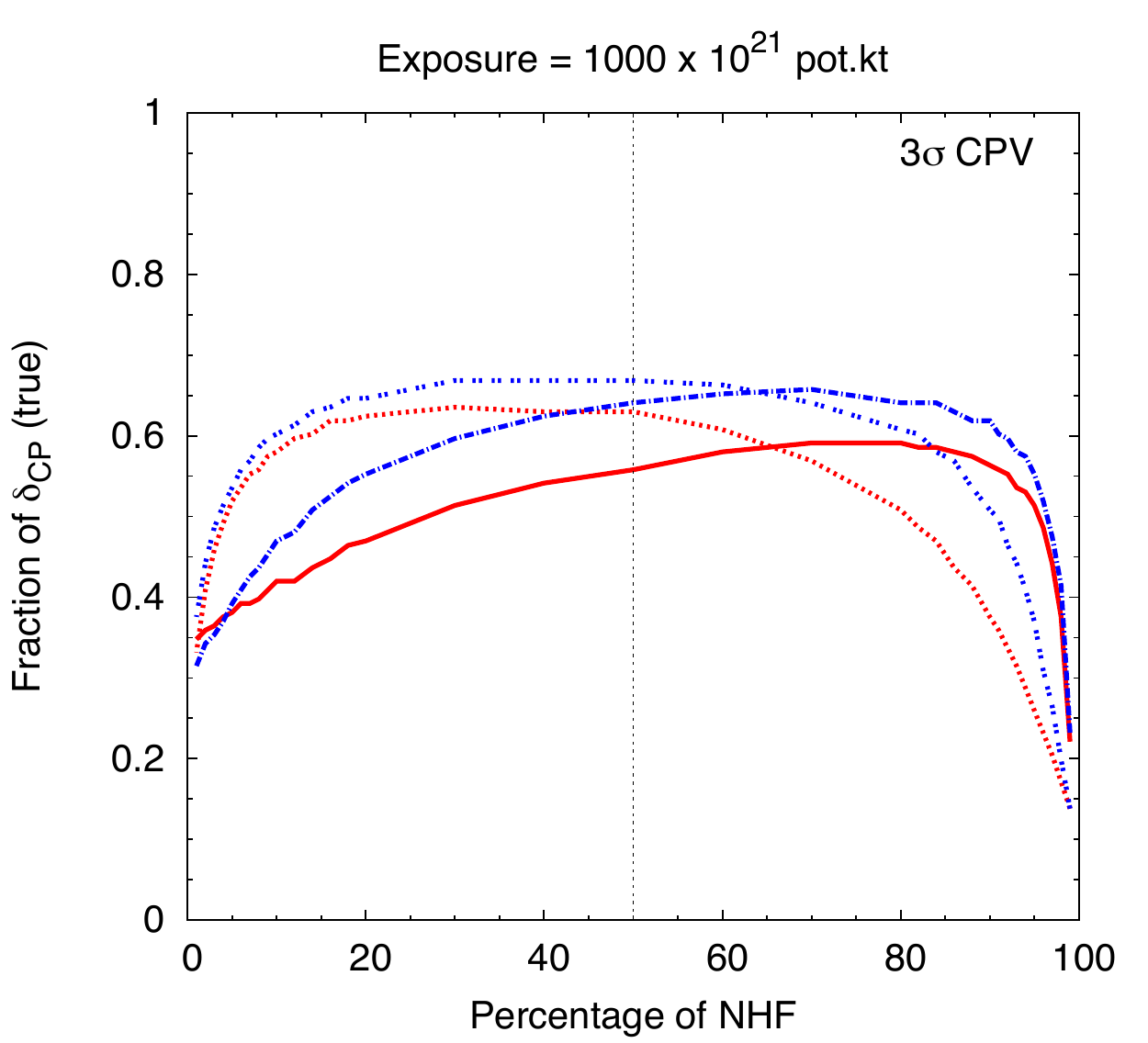}
\includegraphics[width=5.75cm, height=5.75cm]{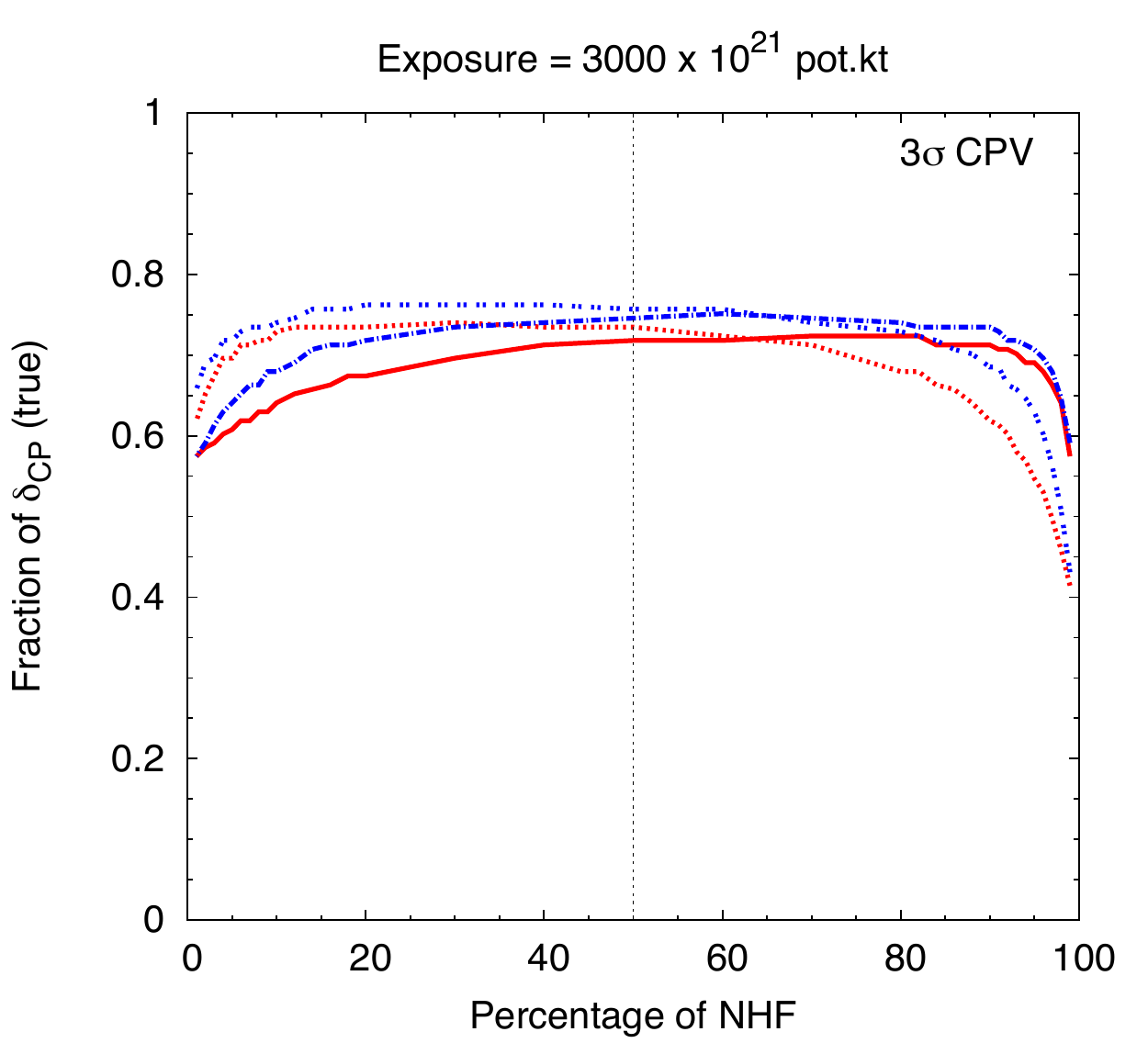}
\mycaption{\label{fig:CPV-ratio} CPV discovery potential at $3\,\sigma$ C.L. as a function of the percentage of NHF in the 
total exposure, for $\stcht = 0.05$. Results are shown for three different benchmark choices of total exposure:    
$300\times10^{21}$ pot$\cdot$kt (left panel), $1000\times10^{21}$ pot$\cdot$kt (middle panel) and $3000\times10^{21}$ pot$\cdot$kt (right panel). 
The results are qualitatively similar for other values of $\stcht$ and confidence levels.}
\end{figure}

In Fig.~\ref{fig:CPV-ratio} we repeat our optimization of the PHF:NHF running, this time with respect to the CP violation discovery potential. Since it is impossible to use $100\%$ coverage as our goal, we instead fix the exposure at values of $300\times10^{21}$ pot$\cdot$kt (left panel), $1000\times10^{21}$ pot$\cdot$kt (middle panel) and $3000\times10^{21}$ pot$\cdot$kt (right panel) - explained shortly - and plot the CP coverage which is obtained when using different ratios of PHF:NHF. Again, we show the results only for $\stcht=0.05$ and $3\,\sigma$ C.L. but have verified that they remain qualitatively similar for other values of $\theta_{13}$(true) and confidence levels: altering $\theta_{13}$(true) effectively just shifts the lines vertically - in the case of smaller $\theta_{13}$(true), the shift is downwards, towards lower CP fractions, and in the case of larger $\theta_{13}$(true) the shift is upwards. We see that a true NH prefers more NHF whereas a true IH prefers more PHF running, which is the opposite to what we found for mass hierarchy discovery. This indicates that whereas the mass hierarchy could reasonably easily be determined simply by accumulating sufficient statistics, in the case of CP violation, the most effective measurement is made when there are roughly equal numbers of $\nu$ and $\bar{\nu}$ events. This is consistent with the intuitive assumption that CP violation is most easily observed by directly comparing $\nu$'s with $\bar{\nu}$'s. In the case of a true NH where the $\nu$ oscillations are enhanced and $\bar{\nu}$ oscillations are suppressed, more NHF is required in order to equalize the number of $\nu$ and $\bar{\nu}$ events. The opposite is true in the case of a true IH. 

In terms of experimental strategy, this means that one should run with an equal mixture of PHF and NHF (shown by the dotted vertical black lines) until the mass hierarchy has been determined. Then, in order to achieve the maximal sensitivity to CP violation in the minimum possible time, if the hierarchy has been determined to be normal, one should continue running with equal PHF and NHF (or with even more NHF if possible) since a NH favors more $\bar{\nu}$ data for CP sensitivity. However, if the hierarchy has been determined to be inverted, one should decrease the amount of NHF to around $30\%$. This applies up until an exposure of roughly $1000\times10^{21}$ pot$\cdot$kt has been reached, after which all admixtures of PHF and NHF with between roughly $20\%$ and $80\%$ NHF give similar results, regardless of the mass hierarchy. Therefore, it is possible to run with just $\sim20\%$ NHF in this regime and still achieve optimal results.

\begin{figure}[t]
\includegraphics[width=5.75cm, height=5.75cm]{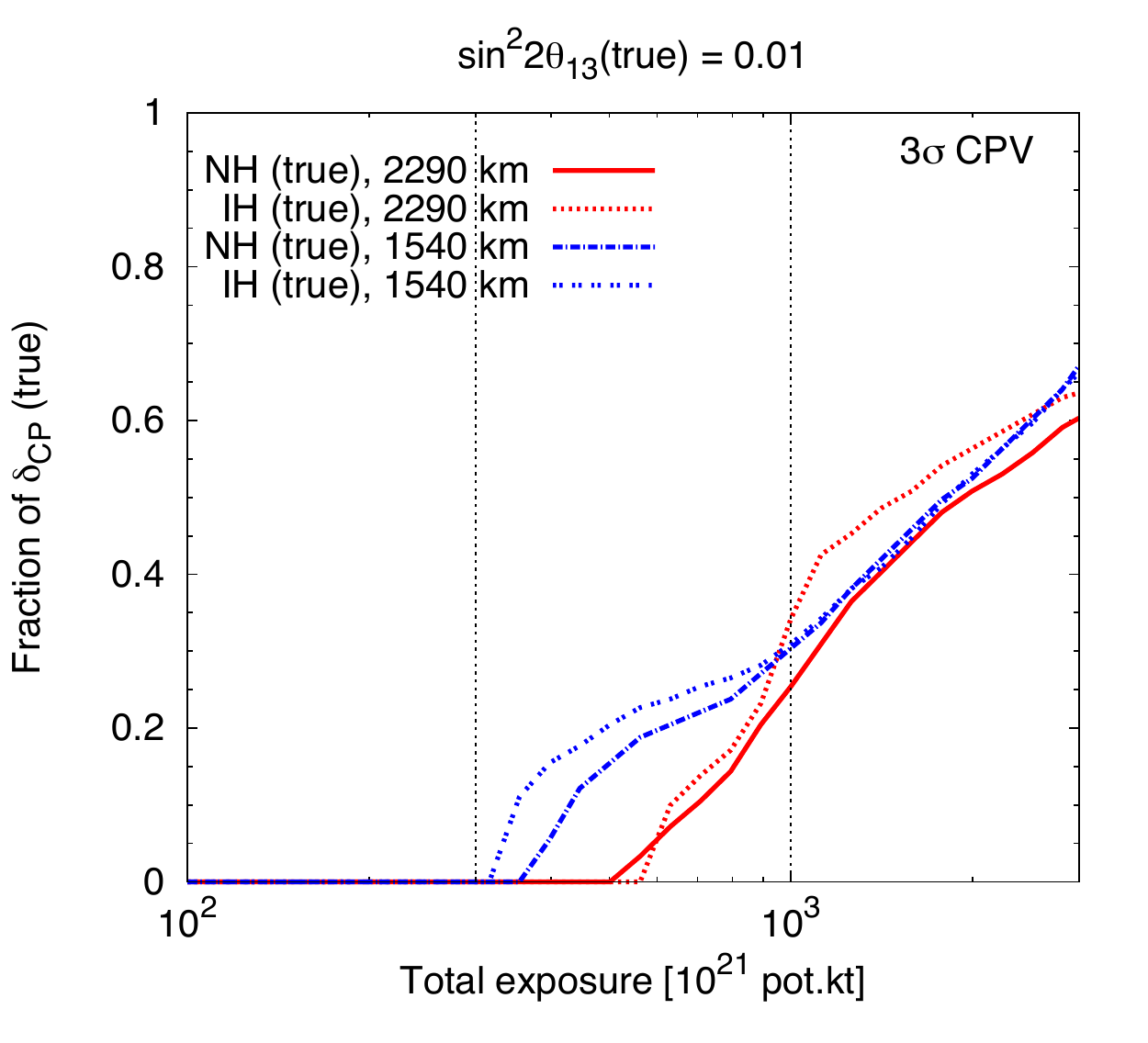}
\includegraphics[width=5.75cm, height=5.75cm]{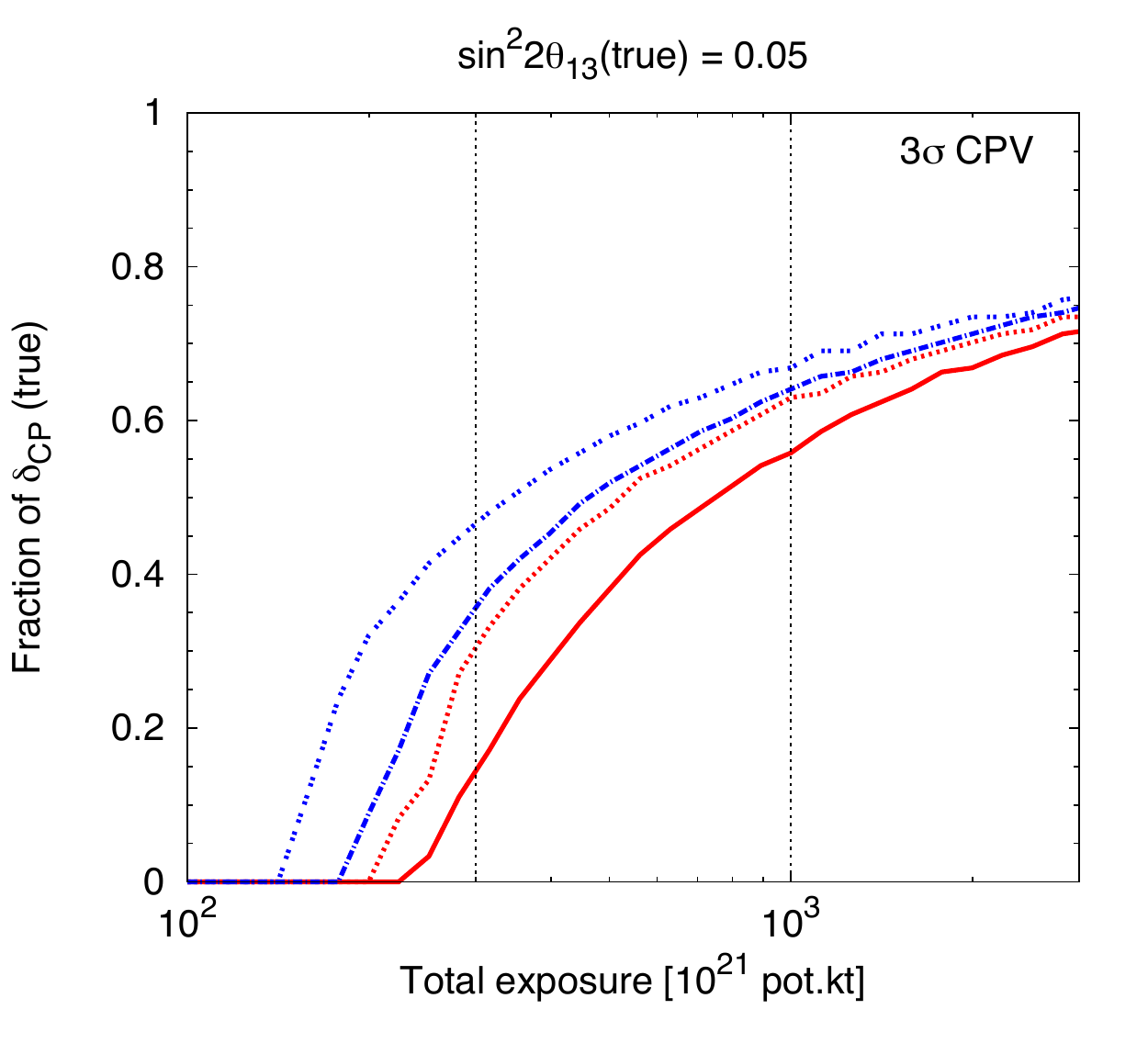}
\includegraphics[width=5.75cm, height=5.75cm]{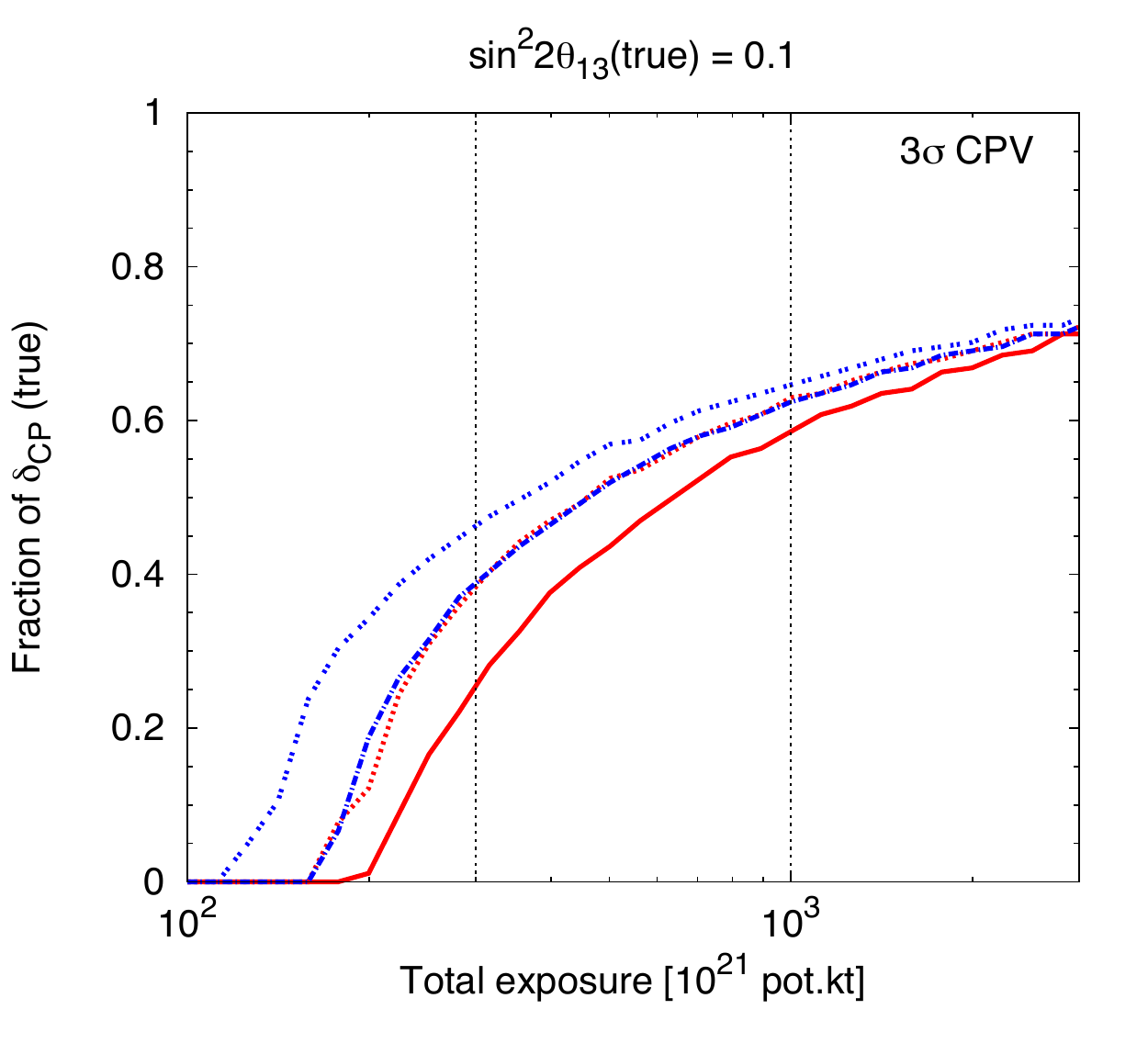}\\
\includegraphics[width=5.75cm, height=5.75cm]{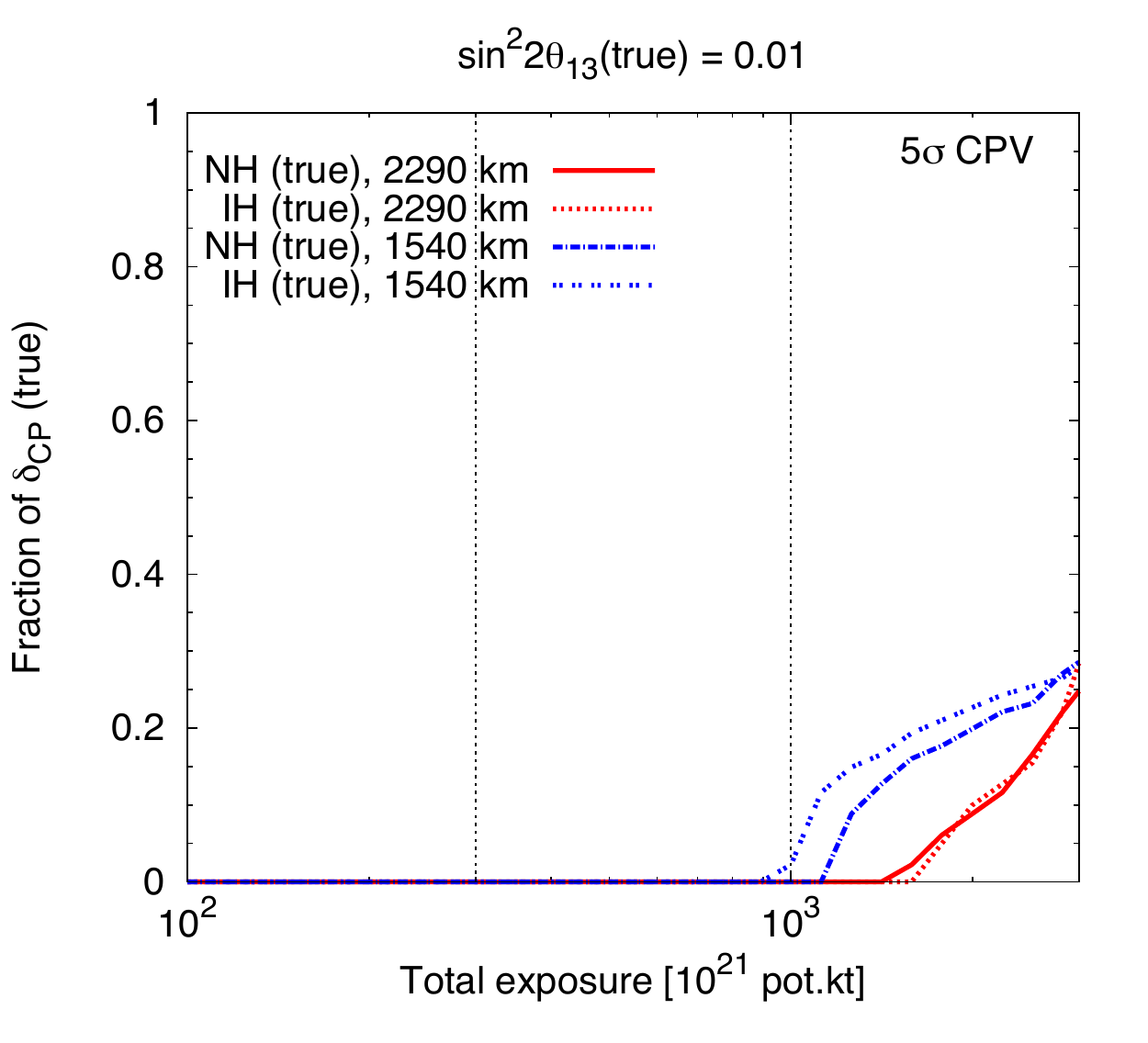}
\includegraphics[width=5.75cm, height=5.75cm]{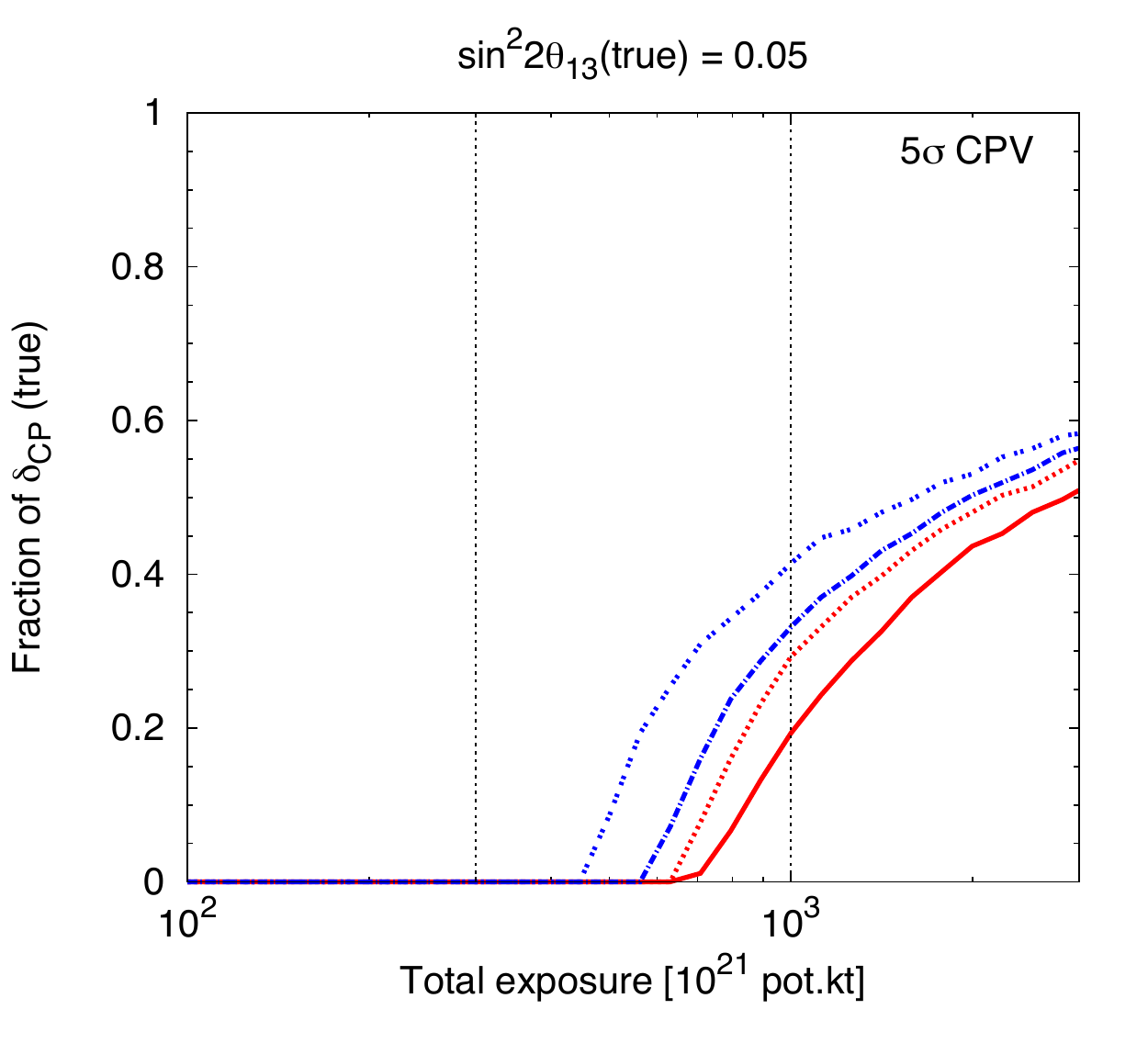}
\includegraphics[width=5.75cm, height=5.75cm]{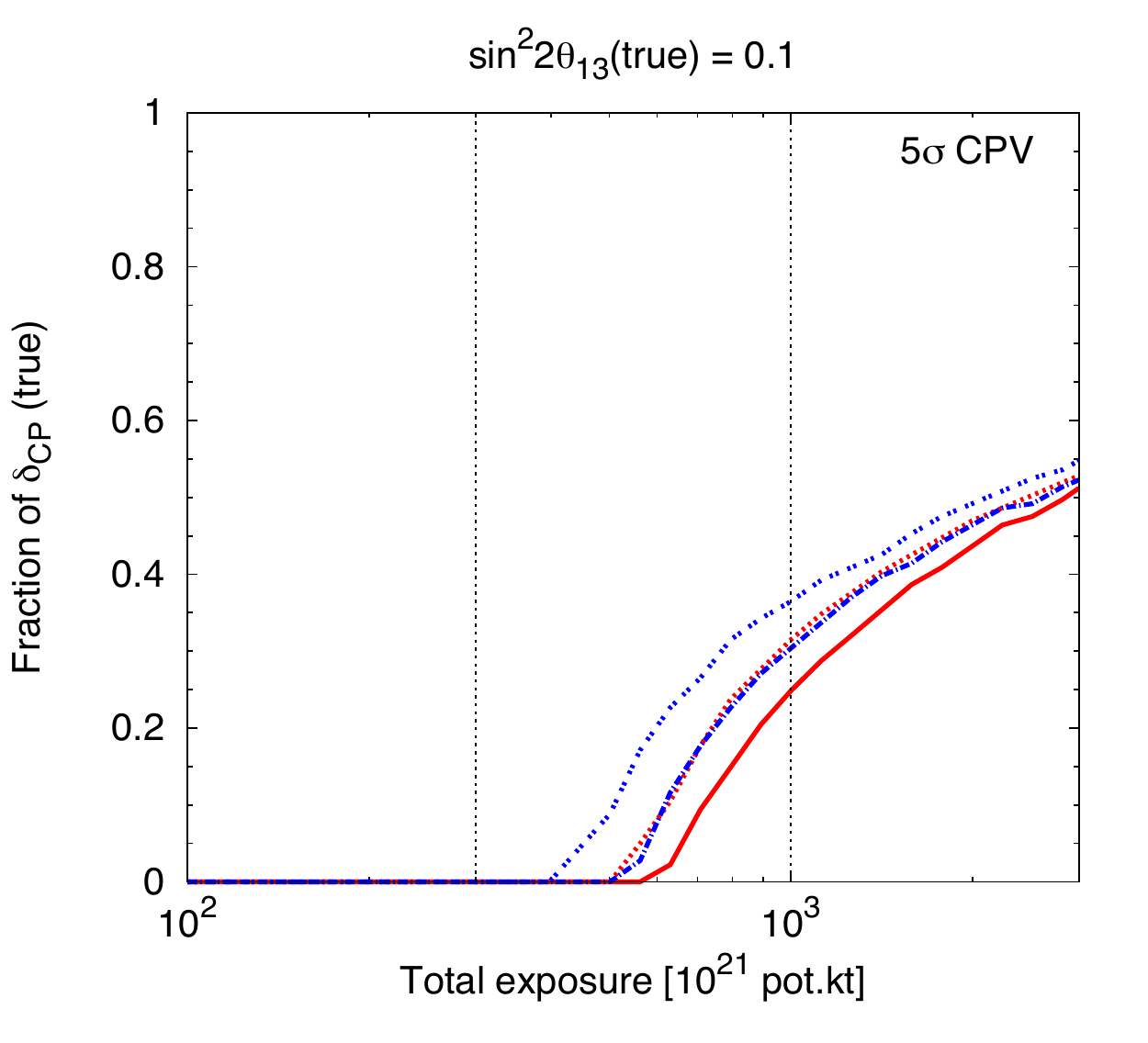}
\mycaption{\label{fig:cpfraction} CP fractions for which a CPV discovery at $3\,\sigma$ (upper panels) and $5\,\sigma$ (lower panels)
confidence levels is possible as a function of total exposure, with equal sharing of PHF and NHF. Results are depicted for three different
choices of $\stcht$: 0.01 (left panel), 0.05 (middle panel) and 0.1 (right panel).} 
\end{figure}

Fig.~\ref{fig:cpfraction} shows the CP sensitivity of the setup as a function of exposure, for $\stcht = 0.01$ (left panel), 0.05 (middle panel), and 
0.1 (right panel), at $3\,\sigma$ C.L. (top row) and $5\,\sigma$ C.L. (bottom row). From the central panel, which corresponds roughly to the current best-fit value of $\stch$, we define three target exposures from which we can achieve three milestone experimental goals of obtaining $3\,\sigma$ sensitivity to CP violation for $30\%,\, 50\%$ and $70\%$ of the $\dcpt$-space. The corresponding exposures are $300\times10^{21}$ pot$\cdot$kt, $1000\times10^{21}$ pot$\cdot$kt and $3000\times10^{21}$ pot$\cdot$kt and are marked by the dotted vertical black lines in all the panels.

In this figure we have considered an equal amount of PHF and NHF, as this is the optimal configuration prior to the discovery of the mass hierarchy as explained previously. In general, for exposures smaller than the maximum of $3000\times10^{21}$ pot$\cdot$kt, the shorter baseline performs better, although the longer baseline only requires slightly more exposure (roughly $40\%$ more) to attain a particular CP fraction. The difference decreases as the exposure is increased and systematic errors become the limiting factor. At the maximum exposure, both baselines perform equally well, yielding a maximum CP fraction at $3\,\sigma$ C.L. of between $60\%$ and $75\%$ for both mass hierarchies, depending on the value of $\theta_{13}$(true). At $5\,\sigma$ C.L. the maximum coverage is $25\%$ to $60\%$, again dependent on the exact value of $\theta_{13}$(true). 

If $\stcht$ is just slightly larger than 0.01, the facility begins to have some $3\,\sigma$ sensitivity to CPV even with our lowest targest exposure. If not, and $\stcht\sim0.1$, our second target exposure of $1000\times10^{21}$ pot$\cdot$kt is sufficient to produce an appreciable $3\,\sigma$ coverage of $\sim30\%$ for both baselines and hierarchies.

\begin{figure}[tp]
\includegraphics[width=5.75cm, height=5.75cm]{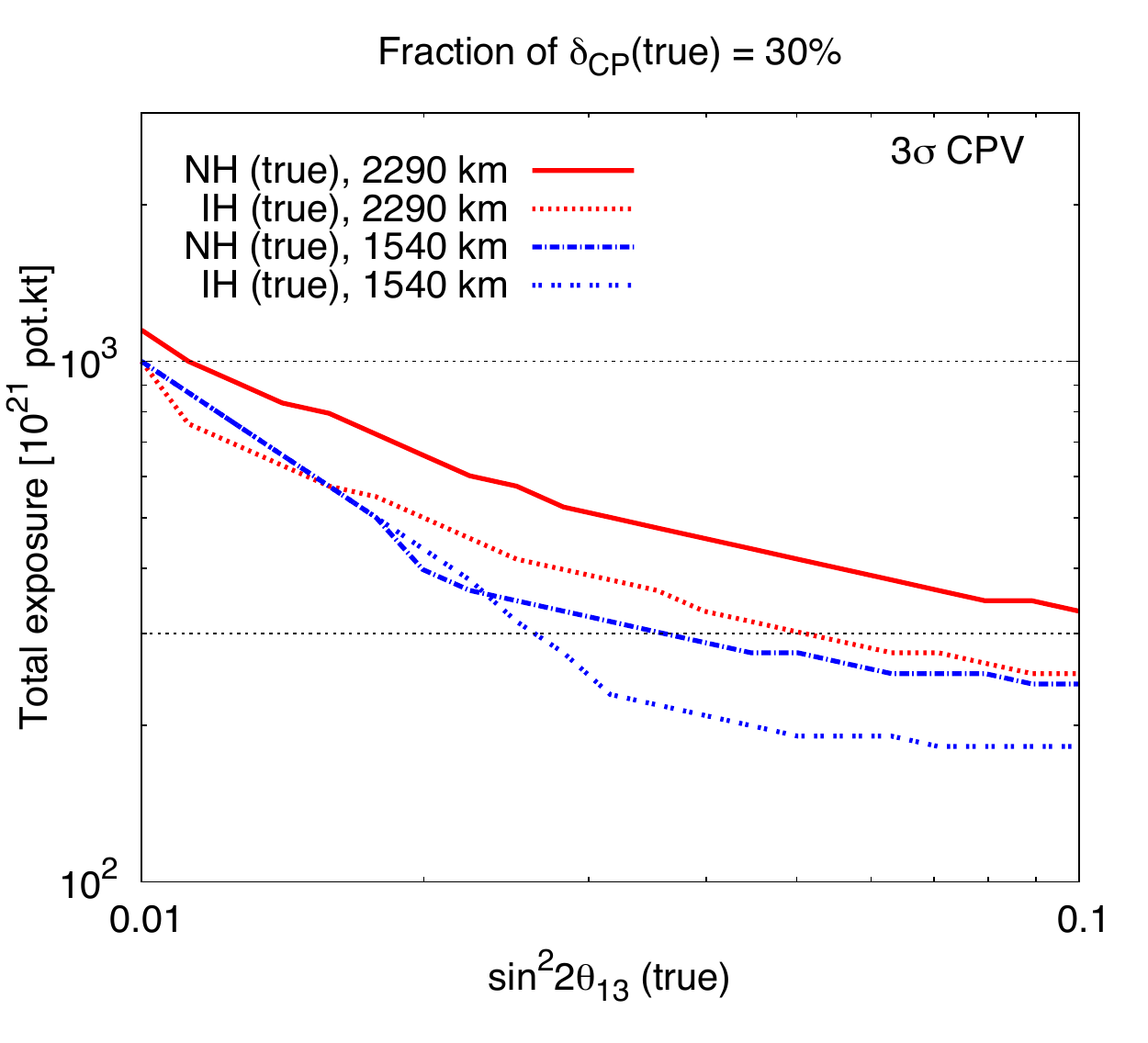}
\includegraphics[width=5.75cm, height=5.75cm]{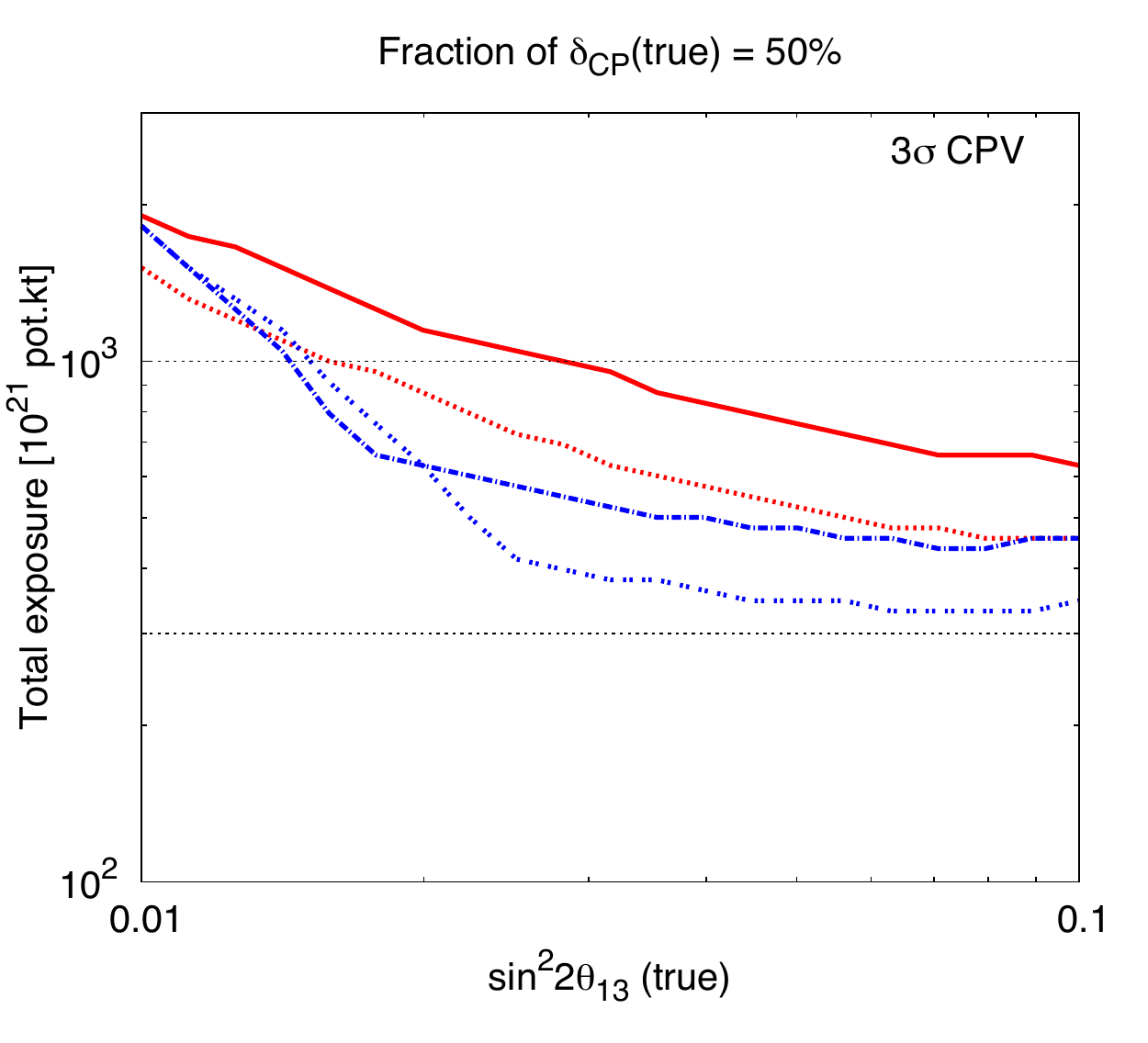}
\includegraphics[width=5.75cm, height=5.75cm]{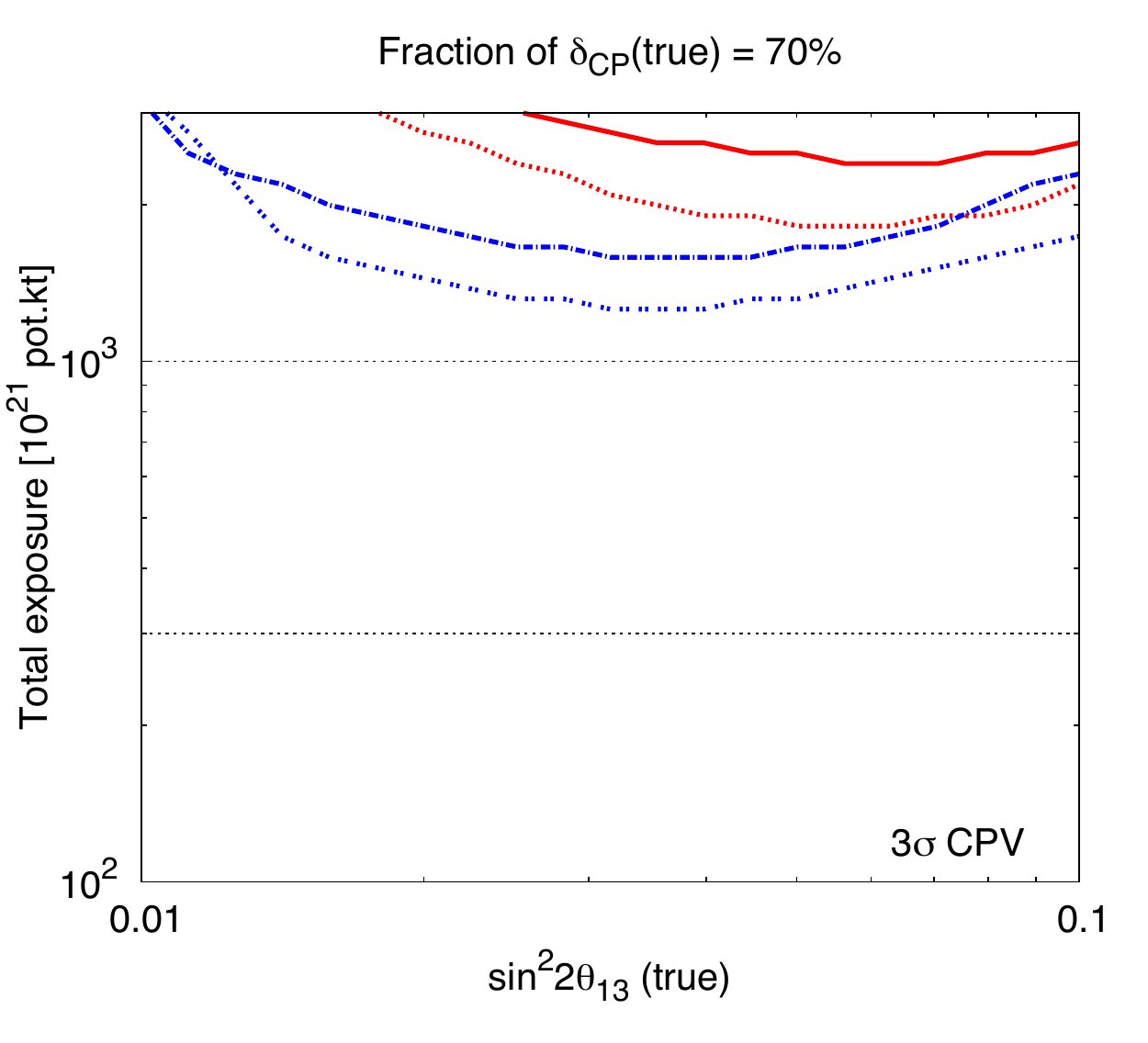}
\mycaption{\label{fig:CPV-exp} Total exposure required as a function of $\stcht$ to have CPV discovery at $3\,\sigma$ C.L. 
for $30\%$ (left panel), $50\%$ (middle panel), and $70\%$ (right panel) values of $\dcpt$. We assume equal sharing of
PHF and NHF in the total exposure.}
\end{figure}

In Fig.~\ref{fig:CPV-exp} we take our three experimental goals - to cover $30\%,\,50\%$ and $70\%$ of the $\dcpt$-space for CP violation - and ask what the minimum exposure is that is required to achieve these goals, as a function of $\stcht$. We have again used equal PHF and NHF but the results do not change significantly if the ratio is altered slightly. For $30\%$ coverage (left panel), we show that our lowest target exposure of $300\times10^{21}$ pot$\cdot$kt (the lower dotted horizontal black line), is that which can provide $30\%$ coverage in the case that $\stcht\gtrsim0.05$ as explained earlier, for both baselines and hierarchies. Our next target exposure, $1000\times10^{21}$ pot$\cdot$kt (the upper dotted horizontal black line), is that which provides $30\%$ coverage if $\stcht=0.01$. In order to reach $50\%$ coverage (middle panel) for this range of $\stcht$, we have to consider exposures up to $2000\times10^{21}$ pot$\cdot$kt (for $\stcht=0.01$). Reaching $70\%$ (right panel) is feasible with our maximum exposure only for certain values of $\theta_{13}$ around 0.05 - if $\theta_{13}$ is either very small or very large, it will not be possible, even with our maximum exposure. Small values of $\theta_{13}$(true) mean that gaining sufficient statistics is a problem, whereas very large values of $\theta_{13}$(true) mean that the atmospheric term of the oscillation probability is so large that it swamps the CP term, effectively acting as a background. However, bearing in mind that the CP coverage which can be obtained by a neutrino factory is $\sim80\%$~\cite{Agarwalla:2010hk}, the 
expected performance of this setup is quite compelling.

\subsection{Precision measurements of $\theta_{13}$ and $\dcp$} 
\label{sec:precision}

\begin{figure}[tp]
\includegraphics[width=5.75cm, height=5.75cm]{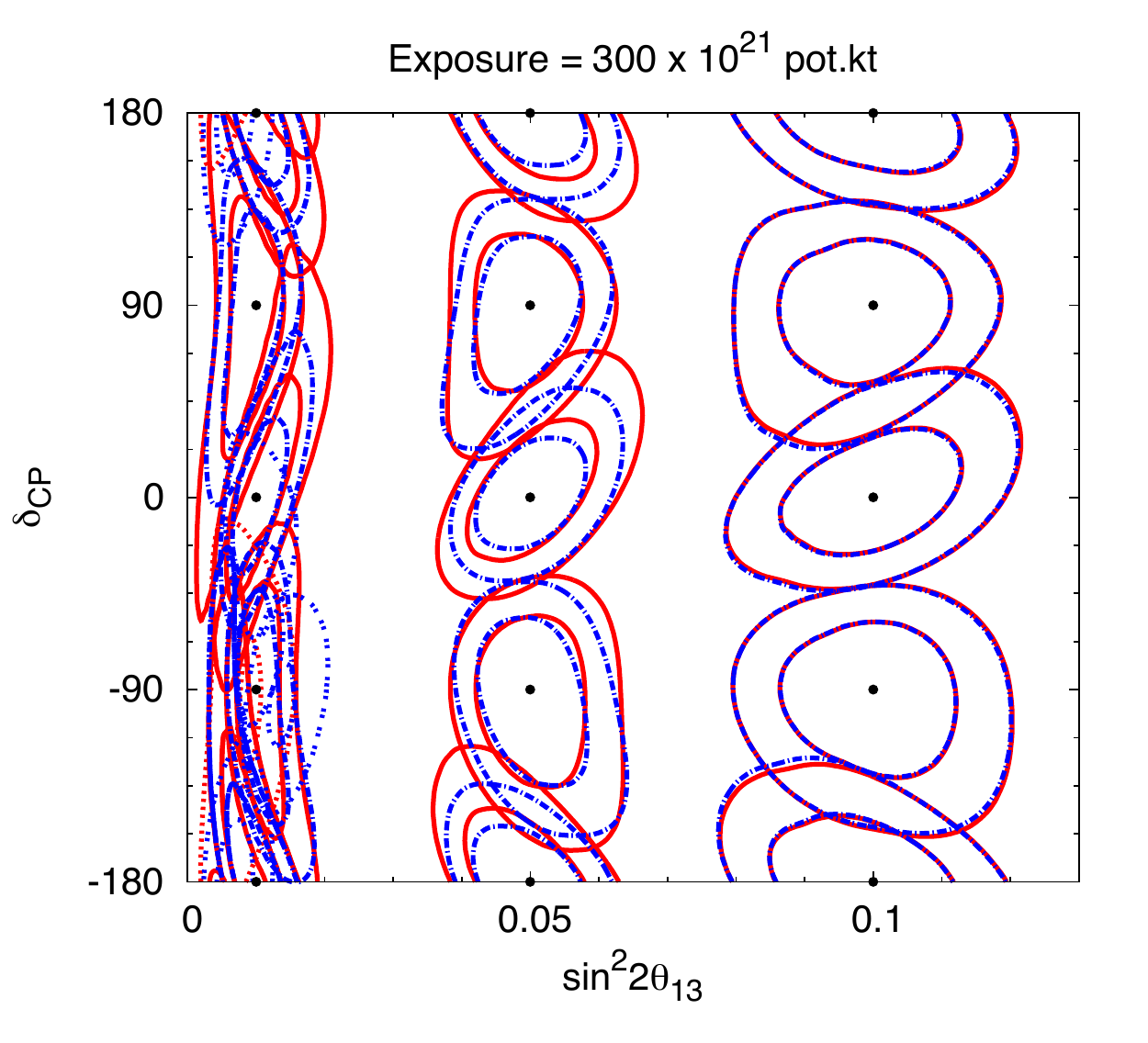}
\includegraphics[width=5.75cm, height=5.75cm]{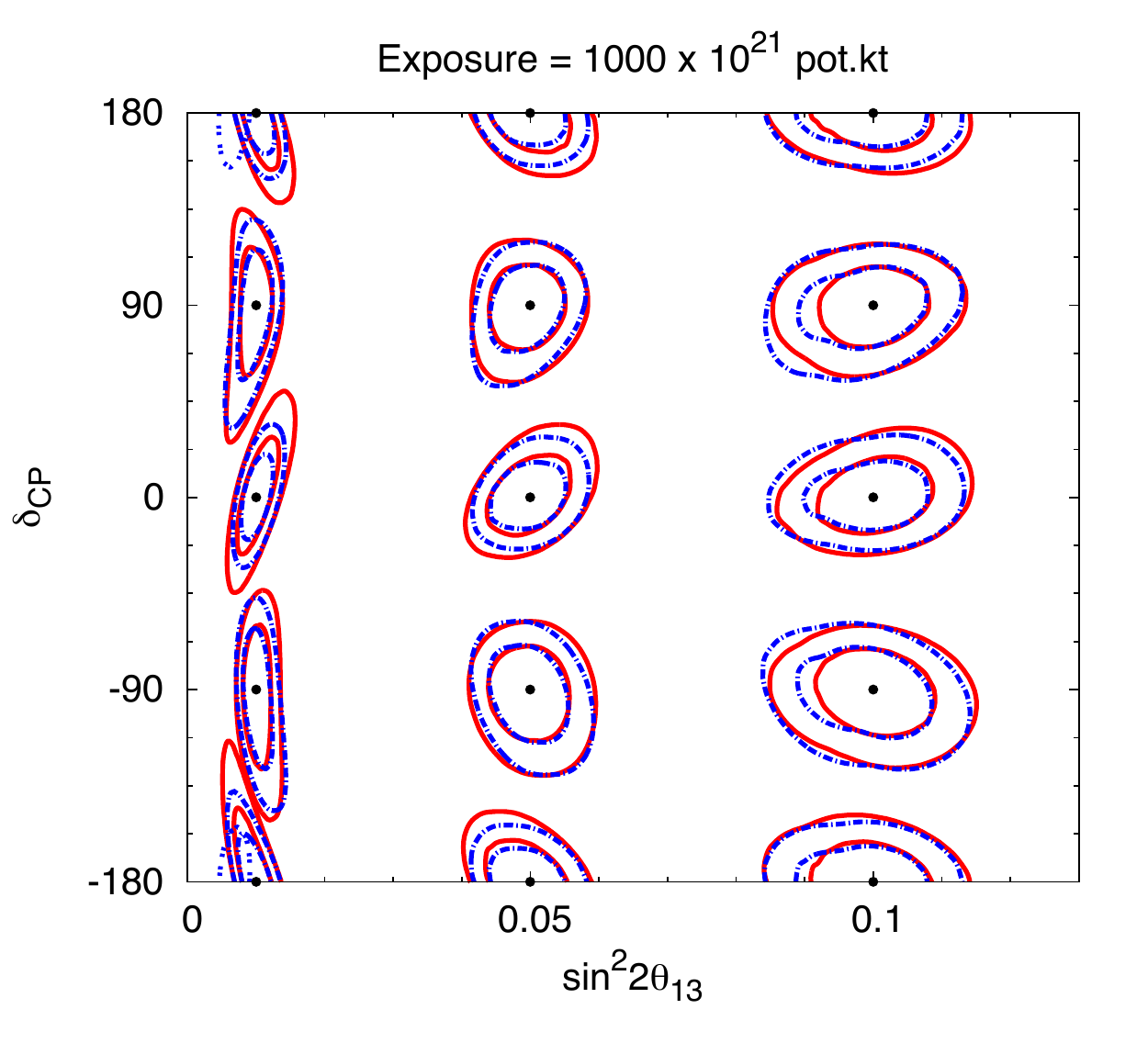}
\includegraphics[width=5.75cm, height=5.75cm]{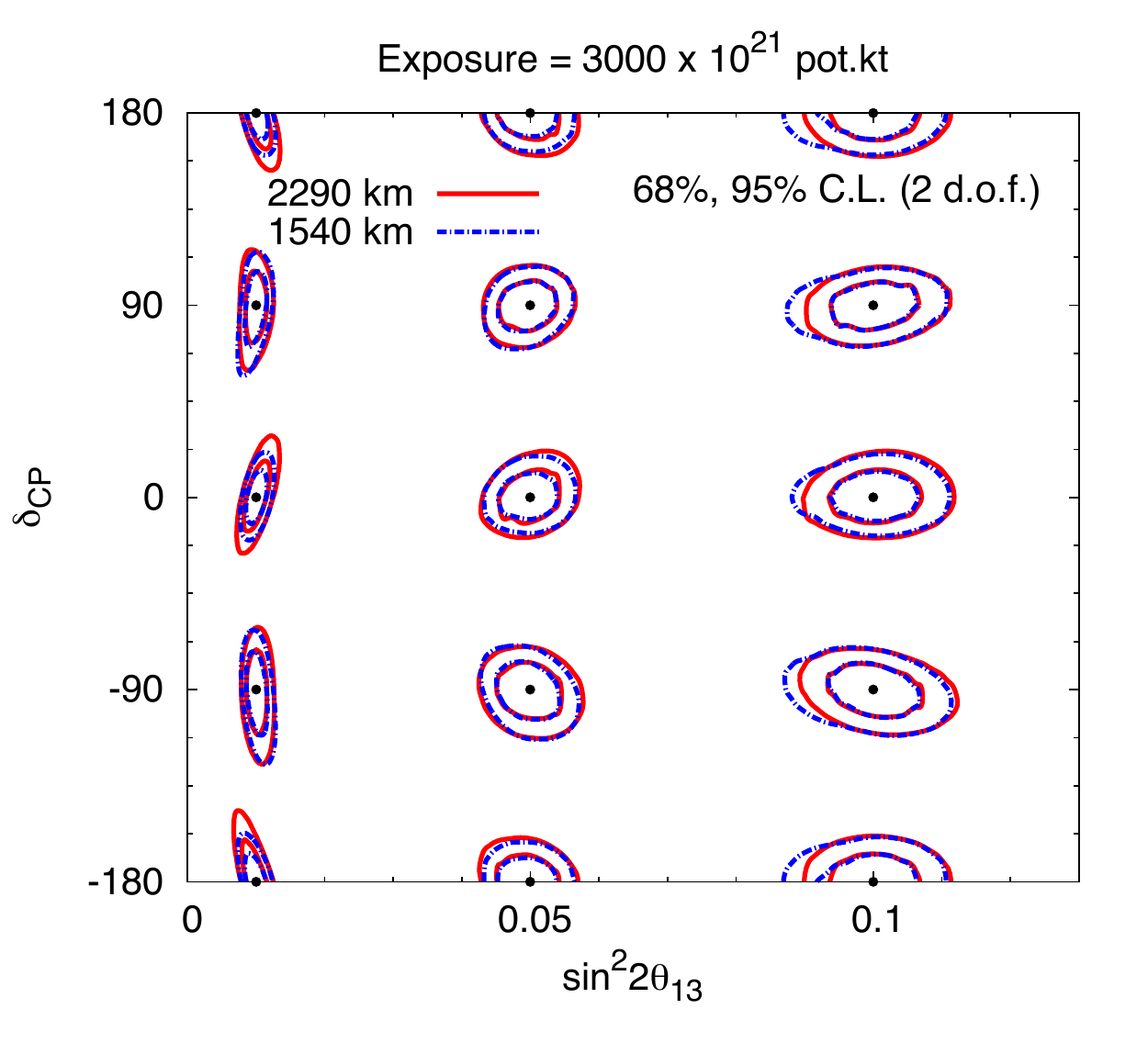}
\mycaption{\label{fig:precision} 68\% and 95\% (2 d.o.f.) C.L. contours in the $\stch$-$\dcp$ plane for 
$\stcht=0.01,\,0.05$ and 0.1 and $\dcpt = 0^{\circ},\,\pm 90^{\circ}$ and $180^{\circ}$ (shown by the black dots). We have simulated a 
true NH; the results for a true IH are qualitatively similar. Equal running of PHF and NHF modes has been assumed with a total exposure of 
$300\times10^{21}$ pot$\cdot$kt (left panel), $1000\times10^{21}$ pot$\cdot$kt (middle panel) and 
$3000\times10^{21}$ pot$\cdot$kt (right panel).} 
\end{figure}

Finally, in Fig.~\ref{fig:precision} we show the fits which can be obtained in the $\stch-\dcp$ plane at 68\% and 95\% C.L. (2 d.o.f). We use our three benchmark exposures of $300\times10^{21}$ pot$\cdot$kt (left panel), $1000\times10^{21}$ pot$\cdot$kt (middle panel) and $3000\times10^{21}$ pot$\cdot$kt (right panel), again with equal sharing of PHF and NHF modes. In each plot, we have simulated true values of $\stcht = 0.01,\, 0.05$ and 0.1 and $\dcpt = 0^{\circ},\, \pm90^{\circ}$ and 180$^{\circ}$ (these true values are marked by the black dots), for both baselines. Here we show the results for the case of a true NH only; the results for a true IH are qualitatively similar.

The dotted contours which are present for $\stcht=0.01$ and the two lowest exposures correspond to the degenerate solutions which can be obtained by fitting to the wrong hierarchy (IH in this case). In other words, for these points, the hierarchy cannot be determined at the $95\%$ C.L.. It is important to note that these degenerate regions do \emph{not} occur at the correct values of $\dcp$ (i.e. they do not occur at $\dcpt$), as do the regions corresponding to the correct hierarchy. For example, the degenerate solution corresponding to ($\stcht=0.01,\dcpt=0^{\circ}$, NH) occurs at ($\stcht=0.01,\dcpt=-90^{\circ}$, IH) so that the true, CP-conserving solution has a degenerate solution at a maximally CP-violating point. Similarly, the degenerate solution for the point $\dcpt=90^{\circ}$ occurs at a CP-conserving point.

We find that when statistics are the limiting factor (small $\stcht$ or low exposure), the $1540\,\mathrm{km}$ baseline produces slightly better results (tighter contours) than 
the $2290\,\mathrm{km}$ baseline, because of the higher flux at the shorter baseline. However, once sufficient statistics have been accumulated, the longer baseline performs as well as, or sometimes slightly better than, the shorter baseline. This is a consequence of the fact that the oscillation peak of the longer baseline covers a wider range of energy bins than the oscillation peak of the shorter baseline (recall that we are using the same energy bins for the analyses of both the baselines) which in turn helps to improve the $\chi^{2}$-fit.

For both baselines, the improvement gained by increasing the exposure is obvious to see - not only do the limits on $\stcht$ and $\dcpt$ decrease, as shown by the shrinking contours, but also the correlations between $\stcht$ and $\dcpt$ are decreased, as shown by the changing shapes of the contours from elliptical (for small $\stcht$ or low exposure) to more circular. In the case that $\stcht=0.01$, our first benchmark exposure will not be sufficient to place any reasonable bound on $\dcpt$, although it can still produce a restriction on $\stcht$. If $\stcht=0.1$, the same exposure is able to define a $95\%$ region of $\sim\pm0.02$ for $\stcht$ and $\sim\pm45^{\circ}$ for $\dcpt$. For our maximum exposure these regions shrink to $\sim\pm0.01$ for $\stcht$ and $\sim\pm20^{\circ}$ for $\dcpt$. We will stress that these results will improve significantly if a prior on $\theta_{13}$ is included, as will all the results presented in this paper.

\section{Summary and conclusions}
\label{sec:conclusion}

\begin{table}[t]
\begin{center}
\begin{tabular}{|c|c|c||c|c||c|c|} \hline\hline
\multirow{4}{*}{Setup}
& \multicolumn{6}{c|}{{\rule[-3mm]{0mm}{8mm}Total exposure ($\times$ $10^{21}$pot$\cdot$kt)}} 
\cr\cline{2-7}
& \multicolumn{2}{c||}{{\rule[-3mm]{0mm}{4mm}$\stcht$}}
& \multicolumn{2}{c||}{\rule[-3mm]{0mm}{4mm}{$\stcht$}}
& \multicolumn{2}{c|}{\rule[-3mm]{0mm}{4mm}{$\stcht$}} \cr
& \multicolumn{2}{c||}{{\rule[0mm]{0mm}{2mm}\bf 0.01}}
& \multicolumn{2}{c||}{{\rule[0mm]{0mm}{2mm}\bf 0.05}}
& \multicolumn{2}{c|}{{\rule[0mm]{0mm}{2mm}\bf 0.1}}
\cr \cline{2-7}
& NH true & IH true & NH true & IH true & NH true & IH true \cr
\hline
{\bf $2290\,\mathrm{km}$} & $1590$ (-) & $1811$ (-) & $32$ ($80$) & $50$ ($158$) & $10$ ($25$) & $20$ ($50$)  \cr
\hline
{\bf $1540\,\mathrm{km}$} & - (-) & - (-) & $158$ ($502$) & $250$ ($627$) & $32$ ($100$) & $50$ ($162$) \cr
\hline\hline
\end{tabular}
\mycaption{\label{tab:mh-compare} Total exposure needed to achieve MH discovery with 100\% coverage in $\dcpt$ for CERN-Pyh{\"a}salmi 
and CERN-Slanic baselines. Results are presented for three different values of $\stcht$ in the range of 0.01 to 0.1. The numbers without (with) parentheses 
correspond to the $3\,\sigma$ ($5\,\sigma$) C.L. results. A dash indicates that this measurement is not possible 
with an exposure below our maximum limit of $3000\times10^{21}$pot$\cdot$kt.}
\end{center}
\end{table}

\begin{table}[t]
\begin{center}
\begin{tabular}{|c|c|c||c|c||c|c|} \hline\hline
\multirow{6}{*}{$\stcht$}
& \multicolumn{6}{c|}{{\rule[-3mm]{0mm}{8mm}Fraction of $\dcpt$ at $3\,\sigma$ (\%)}} 
\cr\cline{2-7}
& \multicolumn{2}{c||}{{\rule[-3mm]{0mm}{4mm}Exposure}}
& \multicolumn{2}{c||}{\rule[-3mm]{0mm}{4mm}{Exposure}}
& \multicolumn{2}{c|}{\rule[-3mm]{0mm}{4mm}{Exposure}} \cr
& \multicolumn{2}{c||}{{\rule[0mm]{0mm}{2mm}300 $\times$ 10$^{21}$ pot$\cdot$kt}}
& \multicolumn{2}{c||}{{\rule[0mm]{0mm}{2mm}1000 $\times$ 10$^{21}$ pot$\cdot$kt}}
& \multicolumn{2}{c|}{{\rule[0mm]{0mm}{2mm}3000 $\times$ 10$^{21}$ pot$\cdot$kt}}
\cr \cline{2-7}
& {\bf $2290\,\mathrm{km}$} &  {\bf $1540\,\mathrm{km}$} & {\bf $2290\,\mathrm{km}$} & {\bf $1540\,\mathrm{km}$} & {\bf $2290\,\mathrm{km}$} & {\bf $1540\,\mathrm{km}$} \cr
\hline
{\bf 0.01} & 0 (0) & 0 (0) & 26 (34) & 30 (31) & 59 (64) & 67 (66) \cr
\hline
{\bf 0.05} & 14 (31) & 36 (47) & 56 (63) & 65 (67) & 72 (74) & 75 (76) \cr
\hline
{\bf 0.1} &  26 (39) & 40 (46) & 59 (63) & 63 (64) & 71 (72) & 72 (73) \cr
\hline\hline
\end{tabular}
\mycaption{\label{tab:compare} Comparison between CERN-Pyh{\"a}salmi and CERN-Slanic baselines for $\dcp$ fractions for which 
a discovery at $3\,\sigma$ C.L. is possible for CPV. Results are shown for three benchmark values of exposure and for three different 
choices of $\stcht$ in the range of 0.01 to 0.1. The numbers without (with) parantheses correspond to the results for a true NH (IH).}
\end{center}
\end{table}

Recent data from neutrino oscillation experiments indicate that the mixing angle $\theta_{13}$ is large, such that $0.01\lesssim\stch \lesssim 0.1$ at the $2\,\sigma$ 
confidence level. Assuming that this hint will be confirmed with high statistical significance in the near future by further running of the T2K experiment, and by data from up-coming experiments, the optimization of next-generation long-baseline neutrino oscillation experiments must incorporate this knowledge. This is in contrast to previous studies which were performed without prior knowledge of $\theta_{13}$, and therefore had to take into consideration all possible values. 

In our study, we have focused on two complementary baselines: $2290\,\mathrm{km}$, corresponding to the CERN to Pyh\"asalmi (Finland) baseline, 
and $1540\,\mathrm{km}$, corresponding to the CERN to Slanic (Romania) baseline. This latter baseline is also close to the FNAL to DUSEL baseline (in the USA) of 1290 km.   
We have considered a CERN-based conventional wide-band beam which provides $3 \times 10^{21}$ protons on target per 
year~\cite{Rubbia:2010fm}, in combination with a Giant Liquid Argon detector. We did not fix the beam running time or the detector mass,
in order to determine how the performances of both the setups vary with exposure (measured in units of pot$\cdot$kt). We thus demonstrated that an
incremental program, staged in terms of the exposure, can achieve the desired physics goals within a realistically feasible timescale, whilst producing new, important and reliable results at each stage. Thereby the risks are minimized, as is the time between the start of the project and the point at which initial results appear. The program can be staged according to future technical developments and as a function of the discoveries achieved at each step, allowing the neutrino beam performance and the far detector mass to be developed in parallel.

A relatively large value of $\tet$ and our choice of a long baseline ensure that it will be possible to determine the neutrino mass hierarchy 
with a `pilot' Liquid Argon detector that has a mass in the range 10 to $20\,\mathrm{kt}$, coupled to a conventional neutrino beam. 
The same setup will test at $>3\,\sigma$ C.L. if CPV is close to maximal and provide, in this case, a first measurement of the CP phase, $\dcp$. 
This approach is fully justified because at present, $\dcp$ angles close to $\pm90^{\circ}$ are not yet excluded, and these are the most easily accessible values.

In the case that $\dcp$ turns out to be inaccessible with the pilot setup, we will have gained the knowledge that $\dcp$ lies in a region 
where we will need more ambitious exposures, achievable by either an increase of far detector mass and/or by a further increase of beam
power. We have found that the reach corresponding to (50\%)\,70\% coverage in $\dcpt$ for CPV will require an exposure of 
$(1000)\,3000\times 10^{21}$ pot$\cdot$kt, ultimately limited by systematic errors. This $\dcpt$ coverage is impressive when compared to
ultimate scenarios, e.g. that of the multi-megawatt power Neutrino Factory which yields a coverage of 75\% to 80\% with a baseline magnetized detector 
of $100\,\mathrm{kt}$.

We found that the performance of the longer baseline, for mass hierarchy determination, is far superior to that of the shorter baseline. At $2290\,\mathrm{km}$ it is possible 
to obtain a $3\,\sigma$ measurement of the mass hierarchy, for all values of $\dcpt$, for the full range $0.01 \lesssim \stcht \lesssim 0.1$. 
The shorter baseline does not perform as well,
especially for the smallest value of $\stcht$ considered. We also find that for $\stcht=0.1$, the shorter baseline requires roughly 3 times more exposure than the longer baseline to make a $3\,\sigma$ determination of the mass hierarchy. For a $5\,\sigma$ measurement the difference is greater. If $\stcht=0.05$ then the shorter baseline requires roughly 5 times as much exposure to make a $3\,\sigma$ or $5\,\sigma$ measurement. Table~\ref{tab:mh-compare} lists the precise exposures required by each of the baselines to achieve 
100\% coverage in $\dcpt$ for mass hierarchy discovery, at the $3\,\sigma$ and $5\,\sigma$ confidence levels, for $\stcht=0.01,\,0.05$ and $0.1$.

In terms of CPV discovery potential, the shorter baseline requires relatively less exposure than the longer baseline to achieve
the same $\dcpt$ coverage, but the difference is not as great as for the hierarchy sensitivity and vanishes rapidly for the larger exposures
and for growing $\stcht$. To obtain a specified $\dcpt$-fraction, the exposure needed for the $2290\,\mathrm{km}$ baseline is at most twice that required for the 
$1540\,\mathrm{km}$ baseline, but the difference is often much smaller. By defining experimental goals to be the discovery of CP violation at the $3\,\sigma$ confidence level, for $30\%,\,50\%$ and $70\%$ of the $\dcpt$ parameter space for the current best-fit value of $\stcht\sim0.05$, we defined three target exposures to define the stages of the experiment: an exposure of $300\times10^{21}$ pot$\cdot$kt is roughly the exposure needed to obtain $30\%$ coverage for both baselines and mass hierarchies, $1000\times10^{21}$ pot$\cdot$kt gives $50\%$ coverage and $3000\times10^{21}$ pot$\cdot$kt provides roughly $70\%$ coverage. Table~\ref{tab:compare} shows the 
CP coverage that can be obtained by each of the baselines for each of these benchmark exposure, at the $3\,\sigma$ and $5\,\sigma$ confidence levels, 
for $\stcht=0.01,\,0.05$ and $0.1$.

Referring back to the hierarchy sensitivity in terms of these benchmark exposures, then with only our lowest exposure, the $2290\,\mathrm{km}$ baseline is 
able to make a $5\,\sigma$ determination of the mass hierarchy for $\stcht>0.05$ for all values of $\dcpt$; the $1540\,\mathrm{km}$ baseline can make a 
$3\,\sigma$ measurement for these values of $\stcht$. 
In other words, a $5\,(3)\,\sigma$ result is \emph{guaranteed} by the 2290 (1540) km baseline with our lowest exposure if $\stcht > 0.05$. 
Tables~\ref{tab:mh-compare} and~\ref{tab:compare} summarize these results quantitatively.

In this work we also quantitatively studied, for the first time, the optimization of the ratio of positive horn focusing (PHF) to negative horn focusing (NHF). Essentially, this is the same as the ratio of neutrino to anti-neutrino data. Firstly, we found that the optimal ratio for determining the mass hierarchy consists of roughly $30\%$ PHF and $70\%$ NHF, for both baselines, bearing in mind that we have to cover both the possibilities of a true normal hierarchy and a true inverted hierarchy. However, taking into account that NHF is experimentally disfavored, we suggest a mixture consisting of equal PHF and NHF for optimal sensitivity to the mass hierarchy. Once the mass hierarchy has been determined to a high confidence level, which will occur at an exposure well below that needed to obtain any significant CP sensitivity, the PHF to NHF ratio can then be altered, depending on the nature of the true hierarchy. A normal hierarchy favors more anti-neutrino data, so the $50:50$ ratio should be maintained, whereas an inverted hierarchy prefers more neutrino data so that the NHF can be reduced to around $30\%$ in this case.

To conclude, we have found that if the current hints of large $\tet$ prove to be true, the measurement of the currently missing neutrino oscillation 
parameters - the neutrino mass ordering and the CP phase $\dcp$ - will be achievable with an experimental program of smaller proportions and 
less ambitious design than those previously envisioned for the scenario of very small $\tet$. To this end, we have demonstrated that a CERN-based 
conventional wide-band beam in conjunction with a Liquid Argon detector at the Pyh\"asalmi mine, at a baseline of $2290\,\mathrm{km}$,
would have exceptional sensitivity to the neutrino mass hierarchy and leptonic CP violation in the light of the emerging hints of large $\tet$.
This setup fits well into an incremental program, staged in terms of the exposure, such that desired physics goals can be achieved within a realistically feasible 
timescale, whilst significant new results are produced at each stage. Staging the exposure is justified in view of the challenges involved 
in operating a high power superbeam and in building a giant underground neutrino detector - a staged approach is therefore effectively safer and 
possibly more cost-effective, whilst also being adaptable to future physics and technical developments.

\subsubsection*{Acknowledgments}

We acknowledge the support from the European Union under the European Commission FP7 Research Infrastructure ÒDesign StudiesÓ 
EUROnu (Grant Agreement No.  212372 FP7-INFRA-2007-1) and LAGUNA (Grant Agreement No. 212343 FP7-INFRA-2007-1).
S.K.A. was in addition supported by the project Consolider-Ingenio CUP.

\bibliographystyle{apsrev}
\bibliography{references}

\end{document}